\documentclass{aastex61}

\usepackage{natbib}
\usepackage{comment}

\newcommand{\Rearth}{\ensuremath{R_{\oplus}}}

\accepted{AJ}

\shorttitle{ASTERIA photometry}
\shortauthors{Knapp et al.}

\begin{document}

\title{Demonstrating high-precision photometry with a CubeSat: ASTERIA observations of 55 Cancri e}

\correspondingauthor{Mary Knapp}
\email{mknapp@mit.edu}

\author[0000-0002-5318-7660]{Mary Knapp}
\affiliation{MIT Haystack Observatory \\
99 Millstone Rd. \\
Westford, MA, 01886}

\author[0000-0002-6892-6948]{Sara Seager}
\affiliation{Department of Physics and  Kavli Institute for Astrophysics and Space Research \\ Massachusetts Institute of Technology \\
77 Massachusetts Avenue \\
Cambridge, MA 02139, USA}
\affiliation{Department of Earth, Atmospheric and Planetary Sciences \\
Massachusetts Institute of Technology \\
77 Massachusetts Avenue \\
Cambridge, MA 02139, USA}
\affiliation{Department of Aeronautics and Astronautics \\
Massachusetts Institute of Technology \\ 
77 Massachusetts Avenue \\
Cambridge, MA 02139, USA}

\author{Brice-Olivier Demory}
\affiliation{University of Bern \\
Center for Space and Habitability \\
Gesellschaftsstrasse 6, \\
3012 Bern, Switzerland}

\author[0000-0002-8781-2743]{Akshata Krishnamurthy}
\affiliation{Jet Propulsion Laboratory,\\ California Institute of Technology\\
4800 Oak Grove Dr.\\
Pasadena, CA 91109}

\author[0000-0003-0103-8820]{Matthew W. Smith}
\affiliation{Jet Propulsion Laboratory,\\ California Institute of Technology\\
4800 Oak Grove Dr.\\
Pasadena, CA 91109}

\author{Christopher M. Pong}
\affiliation{Jet Propulsion Laboratory,\\ California Institute of Technology\\
4800 Oak Grove Dr.\\
Pasadena, CA 91109}

\author[0000-0002-5407-2806]{Vanessa P. Bailey}
\affiliation{Jet Propulsion Laboratory,\\ California Institute of Technology\\
4800 Oak Grove Dr.\\
Pasadena, CA 91109}

\author{Amanda Donner}
\affiliation{Jet Propulsion Laboratory,\\ California Institute of Technology\\
4800 Oak Grove Dr.\\
Pasadena, CA 91109}

\author{Peter Di Pasquale}
\affiliation{Jet Propulsion Laboratory,\\ California Institute of Technology\\
4800 Oak Grove Dr.\\
Pasadena, CA 91109}

\author{Brian Campuzano}
\affiliation{Jet Propulsion Laboratory,\\ California Institute of Technology\\
4800 Oak Grove Dr.\\
Pasadena, CA 91109}

\author{Colin Smith}
\affiliation{Jet Propulsion Laboratory,\\ California Institute of Technology\\
4800 Oak Grove Dr.\\
Pasadena, CA 91109}

\author{Jason Luu}
\affiliation{Jet Propulsion Laboratory,\\ California Institute of Technology\\
4800 Oak Grove Dr.\\
Pasadena, CA 91109}

\author{Alessandra Babuscia}
\affiliation{Jet Propulsion Laboratory,\\ California Institute of Technology\\
4800 Oak Grove Dr.\\
Pasadena, CA 91109}

\author{Robert L. Bocchino, Jr.}
\affiliation{Jet Propulsion Laboratory,\\ California Institute of Technology\\
4800 Oak Grove Dr.\\
Pasadena, CA 91109}

\author{Jessica Loveland}
\affiliation{Jet Propulsion Laboratory,\\ California Institute of Technology\\
4800 Oak Grove Dr.\\
Pasadena, CA 91109}

\author{Cody Colley}
\affiliation{Jet Propulsion Laboratory,\\ California Institute of Technology\\
4800 Oak Grove Dr.\\
Pasadena, CA 91109}

\author{Tobias Gedenk}
\affiliation{MIT Haystack Observatory \\
99 Millstone Rd. \\
Westford, MA, 01886}

\author{Tejas Kulkarni}
\affiliation{Jet Propulsion Laboratory,\\ California Institute of Technology\\
4800 Oak Grove Dr.\\
Pasadena, CA 91109}

\author{Kyle Hughes}
\affiliation{Jet Propulsion Laboratory,\\ California Institute of Technology\\
4800 Oak Grove Dr.\\
Pasadena, CA 91109}

\author{Mary White}
\affiliation{Jet Propulsion Laboratory,\\ California Institute of Technology\\
4800 Oak Grove Dr.\\
Pasadena, CA 91109}

\author{Joel Krajewski}
\affiliation{Jet Propulsion Laboratory,\\ California Institute of Technology\\
4800 Oak Grove Dr.\\
Pasadena, CA 91109}

\author{Lorraine Fesq}
\affiliation{Jet Propulsion Laboratory,\\ California Institute of Technology\\
4800 Oak Grove Dr.\\
Pasadena, CA 91109}



\begin{abstract}

ASTERIA (Arcsecond Space Telescope Enabling Research In Astrophysics) is a 6U CubeSat space telescope (10 cm x 20 cm x 30 cm, 10 kg).  ASTERIA's primary mission objective was demonstrating two key technologies for reducing systematic noise in photometric observations: high-precision pointing control and high-stabilty thermal control.  ASTERIA demonstrated 0.5 arcsecond RMS pointing stability and $\pm$10 milliKelvin thermal control of its camera payload during its primary mission, a significant improvement in pointing and thermal performance compared to other spacecraft in ASTERIA's size and mass class.  ASTERIA launched in August 2017 and deployed from the International Space Station (ISS) November 2017.  During the prime mission (November 2017 -- February 2018) and the first extended mission that followed (March 2018 - May 2018), ASTERIA conducted opportunistic science observations which included collection of photometric data on 55 Cancri, a nearby exoplanetary system with a super-Earth transiting planet.  The 55 Cancri data were reduced using a custom pipeline to correct CMOS detector column-dependent gain variations. A Markov Chain Monte Carlo (MCMC) approach was used to simultaneously detrend the photometry using a simple baseline model and fit a transit model.  ASTERIA made a marginal detection of the known transiting exoplanet 55 Cancri e ($\sim2$~\Rearth), measuring a transit depth of $374\pm170$ ppm. This is the first detection of an exoplanet transit by a CubeSat.  The successful detection of super-Earth 55 Cancri e demonstrates that small, inexpensive spacecraft can deliver high-precision photometric measurements.

\end{abstract}




\section{Introduction} \label{sec:intro}
ASTERIA, the Arcsecond Space Telescope Enabling Research In Astrophysics, is a small spacecraft designed to demonstrate enabling technologies for high-precision space-based photometry from small platforms.  Space-based photometric measurements are a powerful tool for astrophysics, but time on existing large space telescopes is scarce.  Small apertures in space can outperform ground-based telescopes in some metrics, such as temporal coverage and photometric precision.  SmallSats and CubeSats have the potential to increase the availability of precision space-based photometric measurements, but their ability to perform measurements precise enough to be astrophysically useful must be demonstrated before that potential can be fully realized.  The ASTERIA mission was launched to provide such a demonstration by measuring the transit of small exoplanets around nearby stars.

In this paper we describe ASTERIA's photometric performance as demonstrated by observations of the transiting super-Earth 55 Cancri e. Section~\ref{sec:asteria} presents an overview of the ASTERIA mission.  Section~\ref{sec:sciprog} describes the 55 Cancri dataset.  Section~\ref{sec:photometry} describes image processing and transit fitting procedures; Section~\ref{sec:results} presents the retrieved parameters.  We summarize lessons learned from this mission in Section~\ref{sec:discuss}.

\subsection{Exoplanet transit technique}
Out of the thousands of exoplanets and thousands more planet candidates known to orbit main sequence stars, more than three quarters of them have been discovered by the transit technique.  When a planet physically blocks light from its host star as it passes across the star's disk (transit), the star-to-planet area ratio can be measured photometrically as a small drop in the host star's brightness. The power of the transit technique implemented on a space platform is threefold. First, the planet-to-star size ratio is always more favorable than the planet-to-star mass ratio or the planet-to-star flux ratio. Second, transit discovery is not dependent on color or spectra such that a broad bandpass encompassing most of the visible light range can be used, increasing signal. Third, space-based missions above the blurring effects of Earth's atmosphere can reach much higher photometric precision than ground-based telescopes. Ideal satellite orbits do not suffer from the day/night cycle that breaks up transit observations for ground-based telescopes. For a review of space- and ground-based transit surveys see \citet{Deeg2018}. 

\subsection{ASTERIA in context: existing and future transit missions}
Several space-based missions have leveraged these advantages, each with distinct parameter space in terms of star type, star magnitude (and distance), and planet period (Table~\ref{tab:photmssns}). The pioneering Kepler Space Telescope \citep{Borucki2010} discovered thousands of exoplanets transiting Sun-like stars around relatively faint (V=10--15) and distant stars (typically over 1000 light years away).  Nearly all of the Kepler planet host stars are too faint to permit follow-up measurements such as high-precision radial velocity to measure the planet mass. The MIT-led NASA Transiting Exoplanet Survery Satellite (TESS) mission (launched April 2018, \citet{Ricker2014}) is optimized for planets with orbital periods up to two weeks orbiting M dwarf stars; TESS will survey nearly the whole sky in a series of one month observation campaigns, with overlap and correspondingly longer temporal coverage at the ecliptic poles. The proposed ESA Planetary Transits and Oscillations of stars (PLATO) mission \citep{Catala2011} aims to study bright (V of 4--11 mag) stars in a wide field of view (2256 square degrees), with 26 small (12 cm) telescopes mounted on the same platform.  Both TESS and PLATO focus on bright stars amenable to spectroscopic follow-up observations.  Accordingly, these missions use modest aperture sizes (10.5 cm for TESS, 12 cm for PLATO).  

ASTERIA, with a 6 cm aperture, is analogous to a single TESS or PLATO camera.  ASTERIA is a 6U CubeSat technology demonstration mission \citep{Smith2018}. It was originally conceived as a prototype for one element of a constellation of stand-alone small space-based telescopes (Seager et al., in prep.).  Note that the Bright Target Exoplorer (BRITE) constellation \citep{Weiss2014}, while not having high-precision photometric capability, is an early example of a fleet of similar satellites for astronomy.  The Fleet concept, described in Section \ref{sec:nextsteps}, can be thought of as a distributed version of TESS or PLATO, where each camera is part of a free-flying spacecraft.

\subsection{Space-based vs. ground-based transit photometry}
ASTERIA has reached an average photometric precision of 1000 ppm on a 60 second observation of a V$\sim$6 star. For comparison we take HATNet \citep{Bakos2018} a collection of six ground-based telescopes with 11 cm apertures at three different geographic locations. One HATNet telescope can reach a photometric precision of $\sim$3 mmag ($\sim$3000 ppm) in a 3-minute observation for stars at the bright-star end (r$\sim$9.5) \citep{Bakos2018}. ASTERIA, despite a nearly 20 times smaller collecting area, performs slightly better than HATNet.  With respect to larger ($>$ 1 m) ground-based telescopes, the only published 55 Cnc e transit detection from the ground is with the 2.5 m Nordic Optical Telescope (NOT) \citep{DeMooij2014}. The ALFOSC instrument on the NOT reaches a photometric precision of approximately 200 ppm in 7.5 minutes ($\sim$800 ppm in 3 minutes); comparable to ASTERIA's best on the same star (1000 ppm in one minute). \citet{DeMooij2014} determine a star-to-planet radius ratio of 0.0198$^{+0.0013}_{−0.0014}$ 

It may seem remarkable that a tiny space telescope such as ASTERIA has such a high photometric precision as compared to ground-based telescopes. This is because ASTERIA is free from scintillation and other effects of Earth's atmosphere and is therefore at an advantage over ground-based telescopes for observations of bright stars despite its small aperture size. See \citet{Mann2011} for detailed discussion of the challenges of reaching sub-mmag photometric precision with ground-based telescopes.

Scintillation is intensity fluctuation caused when starlight passes through regions of turbulence in Earth's upper atmosphere. Scintillation is seen by the naked eye as stars twinkling. Because scintillation is produced by high-altitude turbulence, the range of angles over which the scintillation is correlated is small, so correction using comparison stars is not usually helpful (but c.f. \citet{Kornilov2012}). 

Estimated scintillation noise for a given star is described by \citet{Young1967} and \citet{Dravins1998} in units of relative flux, 
\begin{equation} \label{eq:scint1}
  \sigma_S = 0.09\ D^{\frac{-2}{3}} \chi^{1.75} (2\ T_{int})^{\frac{-1}{2}}e^{\frac{-h}{h_0}},   
\end{equation}
where $D$ is the diameter of the telescope in centimeters, $\chi$ is the airmass of the observation, $T_{int}$ is the exposure time in seconds, $h$ is the altitude of the observatory in meters, and $h_0$=8000 m is the atmospheric scale height. The constant 0.09 factor in front has a unit of cm$^{2/3}$s$^{1/2}$, such that the scintillation error in units of relative flux. Scintillation does not depend on stellar magnitude and therefore forms a noise floor that limits telescope performance, especially for bright stars.

Using observatory site-specific atmospheric optical turbulence profiles, \citet{Osborn2015} showed that Equation (\ref{eq:scint1}) tends to underestimate the median scintillation noise by a mean factor of 1.5, and provides site-specific correction factors. Attempts to reduce scintillation noise by both scintillation noise correction concepts (e.g.,  \citet{Osborn2015} and observational strategies (e.g., \citet{DeMooij2014} are ongoing. ASTERIA, free from atmospheric effects, focuses on mitigating other sources of non-Gaussian noise that limit photometric performance.

We compare ground and space-based telescopes, assuming only photon and scintillation noise, and an air mass of 1.0 for an integration time of 100 seconds. ASTERIA is comparable to a $\sim$2 m ground-based telescope, with these idealistic parameters.  Here we have estimated fractional precision due to photon noise with
\begin{equation} \label{eq:fracphotprecision}
    \sigma_P = 1/\sqrt{At\eta\Delta\lambda\phi},
\end{equation}
where $A$ is collecting area in meters, $t$ is integration time in seconds, $\eta$ is throughput (here taken to be 30\%), $\Delta\lambda$ is wavelength range (V-band), and $\phi$ is the incident photon flux for a star V=6. The total noise in this idealized comparison is $\sigma = \sqrt{\sigma^2_S + \sigma^2_P}$.

\begin{deluxetable*}{lcccccccc}
\tablecaption{Space-based astronomy photometry-specific missions\label{tab:photmssns}}
\tablewidth{0pt}
\tablehead{
\colhead{Mission } & \colhead{Aperture} & \nocolhead{Photometric} & \colhead{Optimal mag. range}  & \colhead{FOV} & \colhead{Bandpass} & \colhead{Launch} & \colhead{Orbit} & \colhead{Ref} \\
\colhead{Name} & \colhead{[cm]} & \nocolhead{precision} & \colhead{[Vmag]} & \colhead{[deg$^2$]} & \colhead{[nm]} & \colhead{} & \colhead{}  & \nocolhead{}}
\startdata
MOST & 15 & & & 1 & 350--750 & Jun. 2003 & Polar LEO & 1 \\
CoRoT & 27 & & 5.4--16 & 6 & 400--1000 & Dec. 2006 & Polar LEO & 2 \\
Kepler/K2 & 95 & & 8--16 & 116 & 420--900 & Mar. 2009 & Earth trailing & 3,4 \\
BRITE & 3 (x6) & & $\leq$4 & 24x24 & 390--460, 550--700 & 2013, 2014 & Sun sync., LEO & 5 \\
\textbf{ASTERIA} & 6.05 & & $<$6 & 11.2x9.6 & 500--900 & Nov. 2017 & LEO & 6 \\
TESS & 10 (x4) & & 8--13 ($\sim$I band) & 24x24 & 600--1000 & Apr. 2018 & HEO & 7 \\
CHEOPS & 32 & & 6--12 & 0.16 & 400--1000 & 2019 & Sun sync. LEO & 8 \\
PLATO & 12 (x26) & & 4--11 & 1100 (x26) & 500--1000 & 2026 & Sun-Earth L2 & 9 \\
\enddata 
\tablecomments{References: (1) \citet{Walker2003}; (2) \citet{Auvergne2009}; (3) \citet{Borucki2010}; (4) \citet{Howell2014}; (5) \citet{Weiss2014}; (6) \citet{Smith2018}; (7) \citet{Ricker2014}; (8) \citet{Broeg2013}; (9) \citet{Rauer2013}. Note that BRITE, ASTERIA, and CHEOPS focus observations on one specific target star at a time whereas the other missions are surveys for exoplanet discovery.}
\end{deluxetable*}

\section{ASTERIA mission} \label{sec:asteria}
ASTERIA was designed to mitigate two key sources of systematic noise in space-based photometry: time-varying pointing errors and thermal variability.  Errors in spacecraft pointing causes star centroids to drift across pixels on an array detector, inducing systematic variation in retrieved photometry due to intra- and inter-pixel gain variations \citep{Ingalls2012}.  Much like scintillation for ground-based telescopes, pointing error results in a magnitude-independent noise floor for space-based observations.  Thermal variations in the detector, optics, and electronics also induce systematic effects in photometry due to thermally dependent gains, mechanical expansion/contraction, and subtle changes in electronics performance.  ASTERIA has its roots in the 3U CubeSat ExoplanetSat, developed by MIT in collaboration with Draper Lab \citep{Smith2010}.  

\begin{figure}
    \plottwo{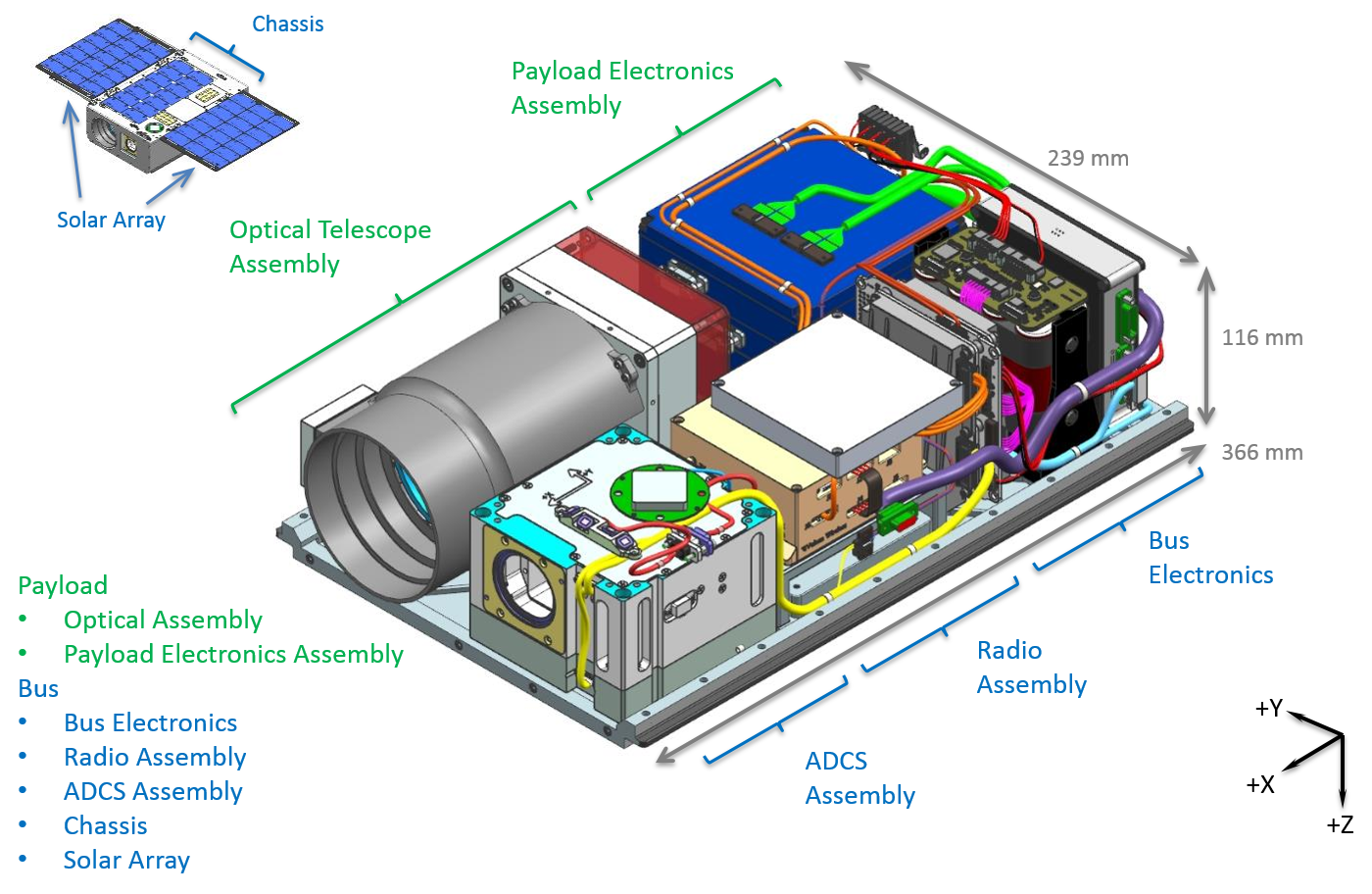}{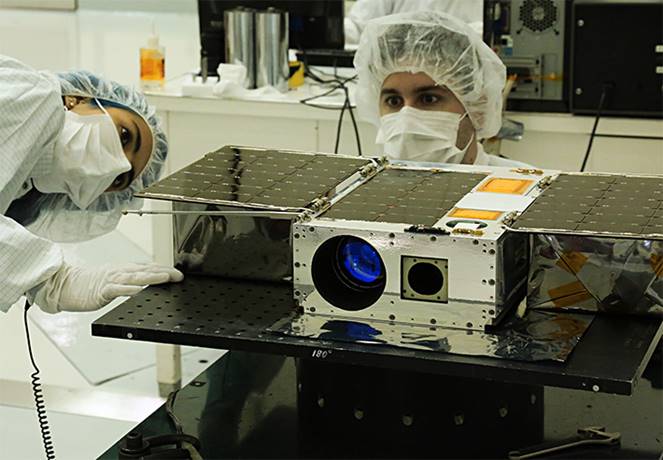}
    \caption{ASTERIA interior layout (left) and fully assembled in the lab (right).  In the interior view, payload components are listed in green; bus components in blue.  Inset at top left shows exterior view, including deployed solar panels.  The photograph on the right shows ASTERIA with solar panels deployed shortly before delivery to the launch provider.  Photo credit: NASA/JPL.}
    \label{fig:insides}
\end{figure}

ASTERIA is a 6U CubeSat with dimensions 239 mm x 116 mm x 366 mm, and mass of 10.2 kg (Figure \ref{fig:insides}).  For a definition of the CubeSat form factor, see \citet{Heidt2000}.  See \citet{Smith2018} for a detailed description of ASTERIA's system and subsystem design, \citet{Bocchino2018} for a detailed description of the flight software framework, \citet{Donner2018} for an overview of mission assurance and fault protection.  
ASTERIA was launched as cargo to the International Space Station (ISS) in August 2017 and deployed into space from ISS on November 20, 2017.  The three month prime mission ended in February 2018 with the successful verification of all technology demonstration requirements.  ASTERIA operations continued through several extended missions for a total of two years (eight times its nominal mission lifetime) until loss of contact in December 2019.

\subsection{Technology demonstration goals}
ASTERIA was a technology demonstration mission rather than a science-driven mission, though the key technologies demonstrated were selected to enable future science missions.  ASTERIA had two key technology demonstration goals: 
\begin{enumerate}
    \item Line-of-sight pointing stability of 5 arcseconds RMS
    \item Thermal stability of $\pm$10 milliKelvin RMS for a single point on the focal plane
\end{enumerate}
ASTERIA achieved 0.5 arcsecond RMS pointing stability on-orbit, approximately 10 times better pointing performance than achieved by other spacecraft in ASTERIA's mass and size category (e.g. BRITE). Fine pointing control was achieved with a two-stage control system.  The coarse attitude of the spacecraft was controlled via reaction wheels and a star tracker.  Residual pointing drift is mitigated by a closed-loop control system that measures centroids of guide stars on the payload detector and actuates a piezoelectric x-y translation stage holding the payload detector.  See \citet{Pong2018} for details of the pointing control system design and on-orbit performance.  

ASTERIA achieved $\pm$5 milliKelvin thermal control on-orbit, measured over a 20-minute period at a single point on the back of the focal plane, approximately 100 times better than achieved by other spacecraft in ASTERIA's mass and size category. Like pointing control, thermal stability is also achieved using a two-stage system.  The spacecraft payload (baffle, optics, focal plane, piezo stage, and readout electronics) is thermally isolated from the rest of the spacecraft using titanium bi-pod mounts and thermally insulating connectors.  Thermal isolation reduces the orbital thermal variation to $\sim$500 milliKelvin amplitude.  A closed-loop control system raises the temperature of the focal plane to a set point slightly above the peak of orbital temperature variation using resistive heaters and holds the set temperature via feedback from high-precision thermal sensors mounted to the focal plane.  See \citet{Smith2018} for thermal control design and performance.

\subsection{Payload and camera design} \label{sect:payload}
ASTERIA's payload is composed of the optical telescope assembly and the payload electronics.  The optical telescope assembly includes a short baffle, a refractive optic (f/1.4, 85 mm), and the CMOS imager mounted to the piezoelectric stage.  A refractive optic was chosen for ASTERIA both for its compactness and for its wide field of view.  The pointing control algorithm needs several bright (V$<$6) guide stars to perform fine pointing adjustments, so a field of view several degrees across was required to ensure that sufficient guide stars were available on the detector for any given pointing.  The telescope was deliberately defocused to oversample the PSF.  The thermal control system is integrated with the optical telescope assembly, which is thermally isolated from the rest of the spacecraft.  See Table~\ref{tab:payloadspecs} for detailed specifications of the optical system and imager.

ASTERIA used a Fairchild 5.5 megapixel CIS2521F CMOS imager as its science detector.  A CMOS imager was chosen over a CCD for several reasons.  The primary factor driving the detector choice is that the imager must be capable of fast readout both to accommodate the 20 Hz fine pointing control loop and to allow unsaturated observation of very bright science targets.  At the time ASTERIA was in development, only CMOS detectors were capable of sufficiently fast readout.  Additionally, the CMOS imager is designed to operate at room temperature, simplifying ASTERIA's thermal design.  See \citet{Magnan2003} for a review of CCD vs CMOS technology.

\begin{deluxetable*}{lc}[h!]
\tablecaption{ASTERIA payload specifications \label{tab:payloadspecs}}
\tablewidth{0pt}
\tablehead{
\colhead{Parameter} & \colhead{Value}}
\startdata
Optics type & Refractive \\
Aperture diameter [mm] & 60.7 \\
Focal length [mm] & 85 \\
Pass band [nm] & 500--900 \\
Lens throughput  & 80\% \\
Detector dimensions [pixels] & 2592 x 2192 \\
Pixel size [$\mu$n] & 6.5 x 6.5 \\
Plate scale [arcsec/pixel] & 15.8 \\
Detector field of view [deg] & 11.2 x 9.6 \\
Quantum efficiency (mean across band) & 42\% \\
Gain [e-/ADU] & 6.44 \\
ADC bit depth & 11 \\
\enddata 
\end{deluxetable*}

The CMOS imager is divided into a top and bottom half; each half has a separate analog amplifier for each column.  The gain of each column's amplifier is slightly different and must be corrected.  There are eight optically dark (physically blocked from light) and eight electrically dark (electrically tied to ground so that photon-induced electrons cannot accumulate) rows at the top and bottom edges of the detector.  The electrically dark pixels are used for bias and flat calibration (Section \ref{sec:cal}).  The detector also has 16 optically dark columns at the left and right edges of the detector for additional calibration; these are not used in the image calibration procedure.

\begin{figure*}[h]
\centering
\includegraphics[width=0.9\textwidth]{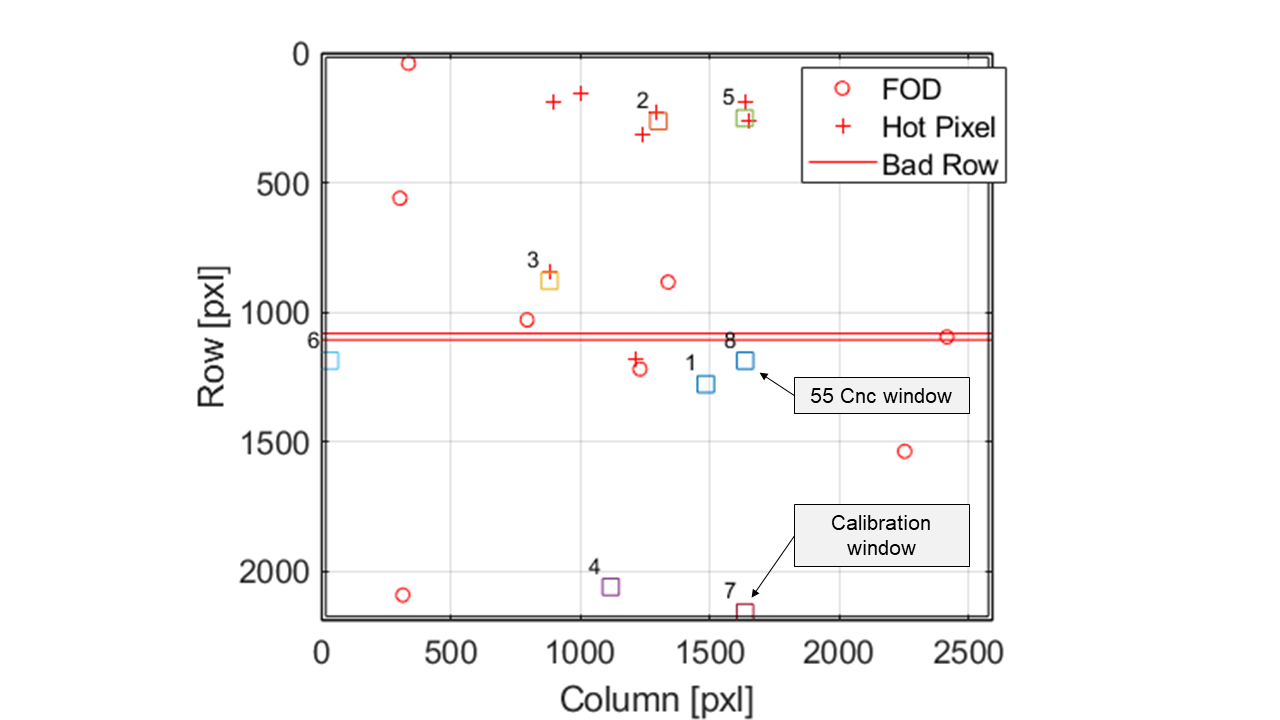}
\caption{Detector layout and window placement for 55 Cnc observations.  Window 8 contains the target star; window 7 is a calibration window placed on the optically and electrically dark pixels at the bottom of the detector (indicated by double line border).  The calibration window covers the same detector columns as the target star window.  The remaining windows are used as guide stars by the fine pointing control system.  The detector is 2592 $\times$ 2192 pixels, corresponding to 11.2$^\circ$ by 9.6$^\circ$ on the sky.}
\label{fig:windows}
\end{figure*}

ASTERIA's CMOS-based camera was customized to read out subarrays of the full imager.  ASTERIA uses this functionality for observations in fine pointing control mode.  Eight 64~$\times$~64 pixel windows, selected and defined on the ground, are centered on bright stars and read out at 20 Hz.  The fine pointing control algorithm calculates centroid positions for each star and calculates the piezo stage motion needed to keep the designated target star motionless on the detector \citep{Pong2012,Smith2018,Pong2018}.  See Figure \ref{fig:windows} for window layout used in 55 Cnc observations. Since the fine pointing control algorithm requires updates at 20 Hz, the maximum integration time in fine pointing mode for both science targets and guide stars is 50 msec.  While some CMOS imagers are capable of reading out subarrays at different integration times, ASTERIA's detector and readout electronics do not allow that mode of operation, so all windows must have the same integration time and be read out at the same time.  The integration time can be set to values less than 50 msec to avoid saturation for bright targets.  

The payload has the capability to sum, or co-add, many 50 msec frames together.  Co-adding significantly reduces data volume while increasing signal-to-noise ratio (SNR) in each image.  In typical operation, 1200 50 msec windowed frames, with eight 64~$\times$~64 pixel windows, are summed into a one minute exposure for each 64~$\times$~64 window.  This one minute cadence data is downlinked to the ground for photometric analysis.  No registration is performed prior to summing because the fine pointing system controls the payload pointing to 0.5 arcseconds RMS, approximately $\frac{1}{30}$th of pixel width (plate scale is 15''/pixel).

\section{Opportunistic science program} \label{sec:sciprog}
The ASTERIA science team selected several target stars of scientific interest that met the needs of the pointing control and thermal control technology demonstrations in order to also demonstrate ASTERIA's photometric capability.  Opportunistic science observations were carried out during ASTERIA's prime mission in parallel with verification of technology demonstration goals.  The demonstration of the fine pointing control system required repeated observations of star fields since guide stars are used to control the piezo stage correction of small pointing errors.  The thermal control system demonstration required the camera system (piezo stage, detector, and payload electronics) to be enabled since those components produce significant thermal loads.  

\subsection{Science targets} \label{sect:scitarg}
ASTERIA's small aperture and maximum integration time of 50 msec limit useful science observations to stars brighter than V=7.  Even brighter stars, V$<$6, were required as guide stars for the fine pointing control demonstration.  Two star fields, \object{55 Cancri} and \object{HD 219134}, were selected for the technology demonstration observations because they provided a favorable set of guide stars and were scientifically interesting targets.  A third field, centered on $\alpha$ Centauri, was observed after the technology demonstration observations had concluded.  Several additional fields were observed during ASTERIA's extended mission, including the eclipsing binary \object{Algol}.  Results from 55 Cancri observations are described in this paper; results from HD 219134 and $\alpha$ Centauri will be presented in future publications.

55 Cancri is a nearby Sun-like star (12.5 pc, spectral type G8 V \citep{vanLeeuwen2007}) that is host to five exoplanets, one of which, 55 Cancri e, is known to transit.  55 Cancri e is a 2\Rearth \ planet with an 18-hour orbital period; its transit was first detected by the Spitzer and MOST space telescopes \citep{Demory2011,Winn2011}.  ASTERIA observed 55 Cancri during the prime mission and first extended mission in an effort to detect the transit of 55 Cancri e and thereby demonstrate a high level of photometric precision.  55 Cancri was the primary target star for the technology demonstration campaign. Table \ref{tab:listobs} lists all observations used to create the final lightcurve described in this paper. All images consisted of 1200x50~ms (60 second) coadded exposures.

\subsection{Observational planning and constraints} \label{sect:observing}
ASTERIA observations must be planned within constraints imposed by both orbital geometry and spacecraft health and safety.  ASTERIA's orbit is similar to that of the International Space Station (ISS), nearly circular with an altitude of $\sim$400 km and orbital inclination of 51.6 degrees to the Earth's equator.  The Earth occupies slightly less than half of the sky visible to ASTERIA.  ASTERIA passes through the Earth's shadow (eclipse) for an average of 30 minutes out of its 92-minute orbit.  Eclipse duration varies  with solar $\beta$ angle throughout the year, with brief periods of no eclipse and a maximum eclipse duration of 35 minutes.  ASTERIA's baffle and thermal control system are designed for operation in eclipse only.

\subsubsection{Observation planning} \label{sect:obs_plan}
ASTERIA observations are planned and sequenced for multiple spacecraft orbits.  An observation is defined as the data collected during one eclipse period.  There is one eclipse per spacecraft orbit, so there is a one-to-one correspondence between observations and orbits.  When multiple observations/orbits are sequenced together, the thermal control system remains active across all observation orbits (both during eclipse and the sunlight portion of the orbit).  Keeping the thermal control system running reduces thermal transients, though the first 1--2 orbits in a sequence exhibit a moderate thermal transient as the control system settles.  Temperature data from sensors on the payload and throughout the spacecraft is available in time-tagged housekeeping data that is downlinked from the spacecraft separately from image data and later written into image FITS file headers.

The camera system is power cycled and reinitialized at the beginning of each orbit just before image data is collected and then again at the end of an orbit after image data collection has finished.  Power cycling and reinitialization of the camera puts the payload in a known good state and reduces the chance of sync loss between the camera FPGA and the flight computer.  Power cycling the camera presents an impulse to the thermal control system since the camera dissipates $\sim$2 W (see \citet{Smith2018}, Figure 27); the overshoot and settling from this impulse is observed in the photometric data (see Figure~\ref{fig:lc_detrend}, top right, Section~\ref{sec:photometry}).  The camera is left on in a free running mode during the sunlit portion of the orbit to maintain a stable thermal environment because of the camera's large thermal dissipation.  When all sequenced observations are complete, the camera and thermal control system are shut off and the spacecraft radio is turned back on.  Image files resulting from the observation are losslessly compressed on board via the standard UNIX tar and gzip algorithms, and downlinked during subsequent communication passes.

\subsubsection{Geometric and orbital constraints} \label{sect:geom_constrain}
Four geometric conditions must be met for viable photometric observations:
\begin{enumerate}
    \item A clear line of sight exists between ASTERIA and the target star (no obstruction by the Earth, Sun, or Moon).
    \item ASTERIA is in Earth eclipse (full umbral shadow).
    \item The payload boresight must form an angle $>$90 degrees with the nadir vector (pointing from the spacecraft to the Earth center).
    \item The payload boresight must form an angle $>$20 degrees with the vector to the Moon.
\end{enumerate}
ASTERIA observes in eclipse only to avoid excess stray light from the Sun\footnote{Test observations have indicated that daylight observations may be feasible in the case where there is a large angle ($>$100 degrees) between the Sun and the target star, but such geometry is not common, so we do not perform daylight observations as part of normal operations.}.  The third constraint on boresight to nadir angle prevents observations close to the Earth limb, where stray light from the illuminated limb degrades photometric data quality.  This is a more restrictive requirement than simple line-of-sight (constraint 1).  The fourth constraint seeks to avoid excess stray light from the Moon.  The geometric constraints listed here are for data quality, not spacecraft safety.  Pre-flight analysis showed that ASTERIA can safely point anywhere in the sky, including at the Sun, though doing so will temporarily reduce the effectiveness of the payload's thermal isolation and the active thermal control system.  

\subsubsection{Operational constraints}  \label{ops_constrain}
Additional constraints are imposed on ASTERIA photometric observations in order to ensure spacecraft safety.  Balancing the power needs of spacecraft subsystems during eclipse observations is critical since all electrical power for the spacecraft is drawn from the storage batteries during eclipse.  The payload and the radio may not operate at the same time because their combined power draw quickly drains the battery.  Furthermore, pre-flight testing revealed interference between the payload and the radio, so the radio must be turned off when the payload is on, and vice versa.  Therefore ASTERIA cannot simultaneously perform observations and communicate with the ground.  

ASTERIA's fault protection system includes protection from battery undervoltage by means of a 43-minute `on-sun timeout' \citep[see Section 3.2.1]{Donner2018}.  Commanded slews to point ASTERIA at a target star must be carefully timed so that the total off-sun time, including slewing to to the target star at the beginning of an observation, eclipse time, and slewing back to a sun-pointed attitude at the end of an observation, does not exceed 43 minutes.  Practically, this means that ASTERIA typically waits until it is already in eclipse to slew to the target star, slightly reducing the available time in eclipse for observations. 

The duration of observations is further limited by the size of the image memory buffer.  Preallocated memory holds co-added images as they accumulate during an observation.  During the prime mission, the hard-coded image memory buffer could hold up to 20 minutes of 1-minute co-added images.  A flight software update at the beginning of ASTERIA's first extended mission increased the image memory buffer allocation to 24 minutes of 1-minute co-added data; a subsequent flight software update increased the capacity to 30 minutes.  Images collected past the image buffer size are simply dropped and not written into memory.

The final operational constraint on observations concerns the buildup of momentum in the spacecraft.  If an attitude maneuver requires the wheels to spin faster than their maximum rate, the wheels are `saturated' and they cannot respond to the requested attitude change.  Torque rods are used to desaturate the reaction wheels by exerting a torque against the Earth's magnetic field, allowing the reaction wheels to decrease their speed.  Observations must be planned such that the momentum does not build up to the point where the reaction wheels become saturated and can no longer hold the commanded star-pointed attitude.  ASTERIA has a residual dipole moment, measured at 0.17 Am$^2$ pre-flight \citep{Pong2018}, which interacts with the Earth's magnetic field and causes momentum to build up quickly in some attitudes.  A MATLAB simulation tool is used to check the momentum build-up for each observation to ensure that the reaction wheels do not become saturated.

Predicting reaction wheel speed during observations is important for data quality as well as spacecraft safety.  Reaction wheels speed changes constantly to control spacecraft attitude and sometimes the speed of one or more wheels will pass through zero, meaning the wheel slows down and reverses its spin direction.  Reaction wheel zero crossings induce a transient in spacecraft pointing (see \citet{Pong2018}, Figure 29), which causes an excursion in star centroid positions on the detector.  We can bias the reaction wheel speeds before an observation in order to avoid zero crossings, though the bias must be carefully chosen to avoid wheel speed saturation.  During the observation planning process, we adjust the wheel speed bias iteratively and examine the resulting wheel speed and momentum predictions in order to choose a bias that avoids wheel speed zero crossings while also keeping accumulated momentum within safe limits. 

\subsubsection{South Atlantic anomaly} \label{sect:saa}
The South Atlantic Anomaly, while not a direct constraint on ASTERIA operations, can cause minor excursions in photometric data.  The South Atlantic Anomaly (SAA) is a region over South America and the South Atlantic ocean where the Earth's radiation belts extend to low altitudes.  Spacecraft in low Earth orbit passing through the SAA experience increased energetic proton flux, which can cause single event upsets in electronics and hot pixels in array detectors.  ASTERIA does operate and observe through the SAA.  Data collected at 20 Hz from the pointing control system indicate that transient hot pixels appear on the detector during SAA passages \citep[Figure 25]{Pong2018}.  These transient hot pixels last for one or two 50 msec frames before returning to background levels.  The pointing disruptions caused by transient hot pixels are short compared to 1-minute co-added exposures and do not affect photometric precision \citep[Figure 28]{Pong2018}.  No build-up of persistent hot pixels has been observed.  

\section{Photometric data analysis} \label{sec:photometry}
Data from ASTERIA observations are downlinked as binary files and then translated into FITS files using custom python software.  Metadata, including spacecraft housekeeping telemetry (e.g. temperature measurements), are included in the FITS header for detrending.

The primary data reduction challenge for ASTERIA is addressing column-dependent gain variation (CDGV). CDGV is specific to CMOS detectors because each column is tied to a specific amplifier and analog-to-digital (ADC) converter. Additionally, each pixel has its own amplifier. We performed extensive characterization of the detectors in the laboratory and tested various reduction recipes on the simulated data in order to find the optimal solution for ASTERIA data, which is presented below.  For details on the laboratory characterization and reduction strategies assessed, see \citep{Krishnamurthy2020}.

\subsection{Calibration data} \label{sec:calib_data}
Traditional calibration frames (biases, darks, flats) were not collected during ground testing because they were not required to verify ASTERIA's technology demonstration requirements.  Approximations for these calibration frames were collected in flight with mixed success.

Bias (zero exposure time) and flat (uniform illumination) frames are typically used to correct for offsets and pixel-to-pixel gain variation, respectively.  Dark frames are not needed for ASTERIA data reduction due to the short 50 msec integration time.  Near-zero exposure time bias frames were collected on orbit in windowed mode at the minimum integration time the detector can support (22.6 $\mu$sec).  Individual bias frames are coadded onboard the spacecraft in the same manner as light images.  Bias frames collected in this manner (i.e. at a different time than the light images they are used to correct) were found to be less effective than calibration windows collected simultaneously with the target star images.  The calibration window was placed at the upper or lower edge of the detector to capture the electrically/optically dark pixels.  The calibration window covered the same detector columns as the light image.  See Figure~\ref{fig:windows}, where window 7 is the calibration window for window 8.  

Flat frames were approximated by using stray sunlight to illuminate the detector.  The MOST space telescope used stray light illumination for on-orbit flat frames, although the stray light illumination was an unintended effect \citep{Rowe2006}, while the Hubble Space Telescope has used the illuminated Earth for flat fielding \citep{Holtzman1995}.  In order to capture stray light flats, ASTERIA was commanded to an attitude with an offset angle of 40 degrees from the sun vector during orbit day.  The spacecraft then rotated about the camera boresight vector while collecting images to smear out the stray light illumination.  Flat frames were also summed onboard the spacecraft to increase signal. This method produced images with 25-50\% illumination as desired, but the illumination was not uniform across the detector.  

\subsection{Data reduction} \label{sec:cal}
We begin the data reduction process by manually reviewing every coadded image and discarding frames that were obtained during the orbital ``sunrise'' or ``sunset.'' These frame appear either at the beginning or end of the orbit, and display significant background contamination from stray light.  In a few cases, a bug in the pointing system caused the stars to be offset from the center of their windows; these data were discarded as well.

We then proceed to bias and background correction. The median of the electrically dark bias column from the calibration window is subtracted from all pixels in the respective column in the science frame.  We then subtract background/sky noise by selecting pixels in the target star window (rows 0-15) that are devoid of any stellar flux, taking the median of the pixels for each column, and subtracting that median value from every pixel in that column. 

Finally, we correct for column-dependent multiplicative gain. As noted in \ref{sec:calib_data}, the on-sky flatfield image was not evenly illuminated in region of the detector corresponding to the 55 Cnc window, and so could not be used for this purpose.  Instead, for the 55 Cnc dataset, we divide each pixel in a given column by the normalized median of the corresponding bias column from the calibration window. 

\begin{figure*}[h]
\centering
\includegraphics[width=0.9\textwidth]{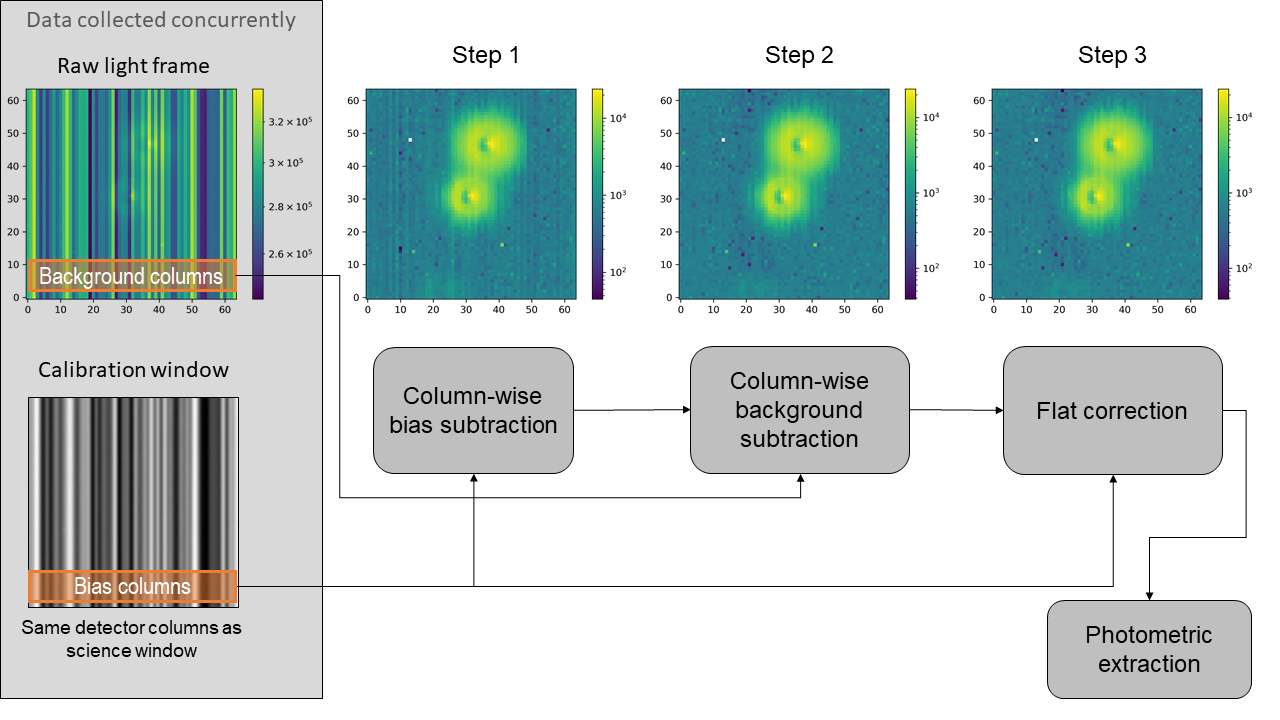}
\caption{55 Cnc image calibration steps.  The star closest to the center of the window is 55 Cancri; the star to the upper right is \object{53 Cancri}.  The frame is composed of 1200 50 msec exposures summed onboard ASTERIA.  The window is 64x64 pixels square.  See Figure \ref{fig:windows} for the locations of the light and calibration windows on the detector.}
\label{fig:comparewincorr}
\end{figure*}

The output of each processing step is shown in Figure~\ref{fig:comparewincorr}. The target star is visually cleaner than the raw image, although there is still some residual column-dependent noise.

\begin{figure*}[h!t]
\begin{center}
\includegraphics[width=6.0in]{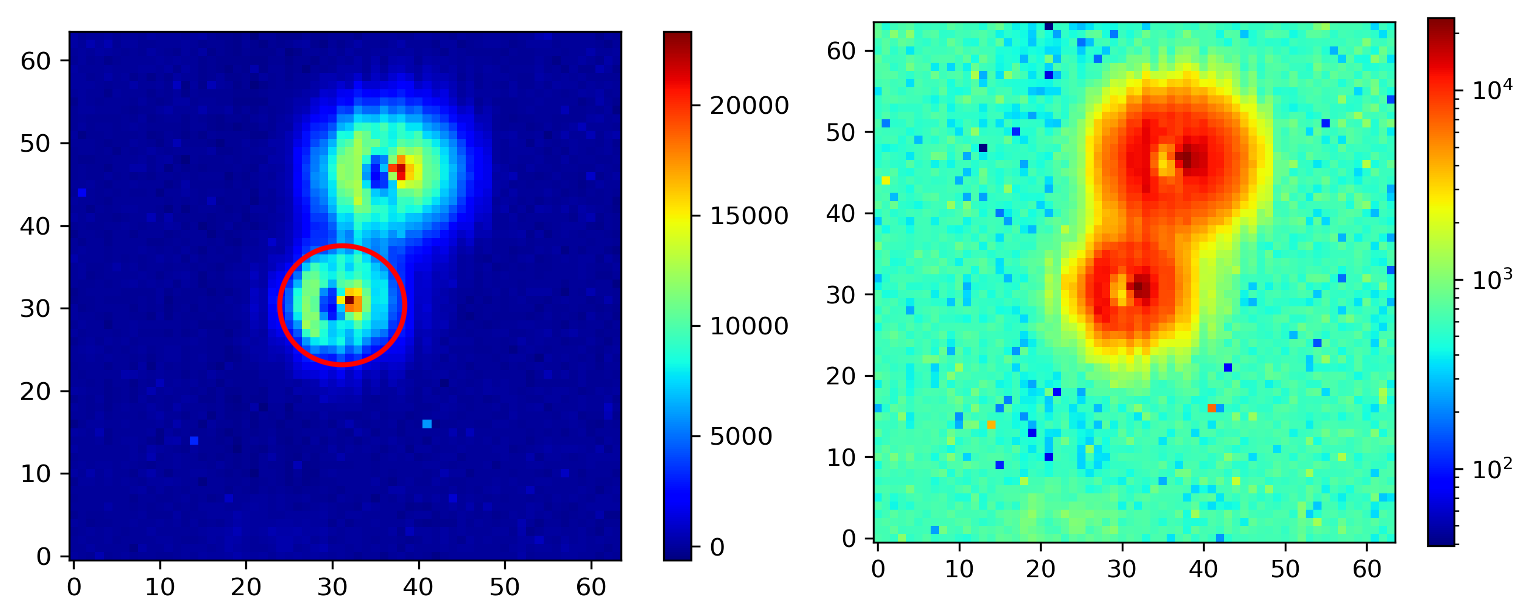}
\caption{(Left) Stellar image after performing removal of bias and background, and correction for flat field variations.  The red circle indicates the photometric aperture. (Right) Reduced image in the log scale still showing column-dependent gain variations especially around the target. Note that 53 Cnc looks larger despite being slightly fainter than 55 Cnc in V band (V=6.23, \citet{Ducati2002}) due to its redder spectrum; chromatic aberration in the ASTERIA optics produces a larger PSF at longer wavelengths.}
\label{fi:final}
\end{center}
\end{figure*}

\subsection{Centroiding and aperture photometry} \label{sec:extract}
In order to measure the target star x/y position on the detector, we compute the flux-weighted 1st moments for each pixel in the subarray window along x and y. We use the mean centroid position to determine the target center, which is used as input to the aperture photometry procedure. We compute the fluxes using an array of apertures with radii ranging from 5 to 13 pixels. Observations of 55 Cnc are complicated by the spatial proximity of \object{53 Cnc} (about 20 pixels), which is $\delta$V=0.28 fainter and partially blended with 55 Cnc.  We compute for each aperture size the photometric precision given by the RMS of the light curve. As the flux contamination from 53 Cnc increases with aperture size, the contribution from background noise overtakes the improvement in photometric precision. An optimal aperture with a radius of 10 pixels is obtained based on the minimization of RMS. We use the \texttt{CircularAperture} routine from the \texttt{photutils} \citep{Bradley_2019_2533376} python package to perform the photometric flux extraction from the subarray windows using a fixed hard-edged aperture with radius of 10 pixels for all 55 Cnc data.

Visual inspection of the raw time-series shows that each visit starts with a telescope settling phase characterized by marked increased of the PSF FWHM for all stars of $\sim$20\%. This effect is likely due the thermal control system overcompensating for the drop in thermal load when the camera briefly turns off during reinitialization before the observation begins. We find however that the centroid positions and the overall background level are not affected.

The proximity of 53 Cnc to 55 Cnc in ASTERIA's subarray aperture motivates us to attempt PSF fitting to extract the photometry. However, the ASTERIA PSF is field-dependent; guide stars from other windows cannot serve as PSF references. Thus, the only available PSF reference for 55 Cnc was 53 Cnc. Additionally, chromatic aberrations in ASTERIA's optics broaden the PSF of the redder 53 Cnc relative to 55 Cnc. We therefore find that aperture photometry outperforms PSF fitting.

\begin{figure*}[h]
    \gridline{\fig{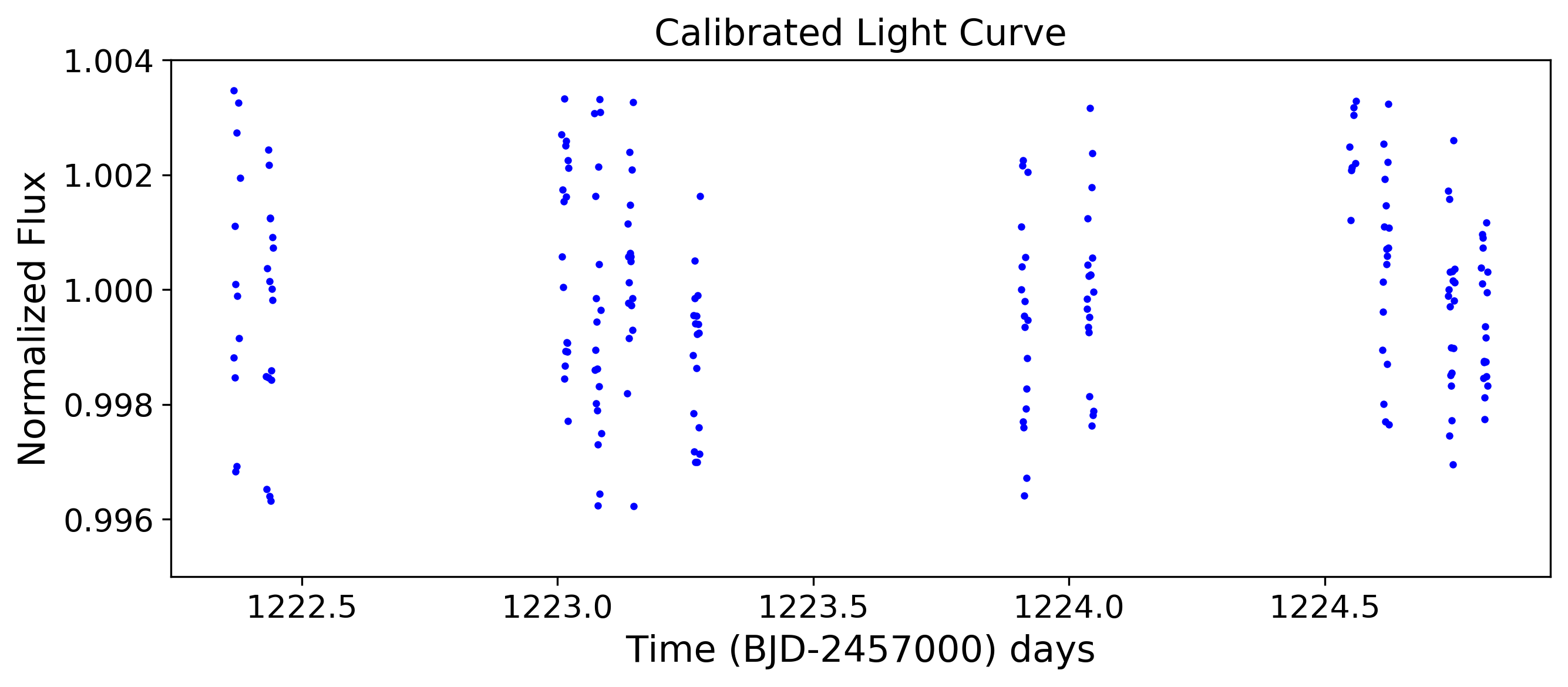}{0.45\textwidth}{Raw lightcurve after image calibration}
              \fig{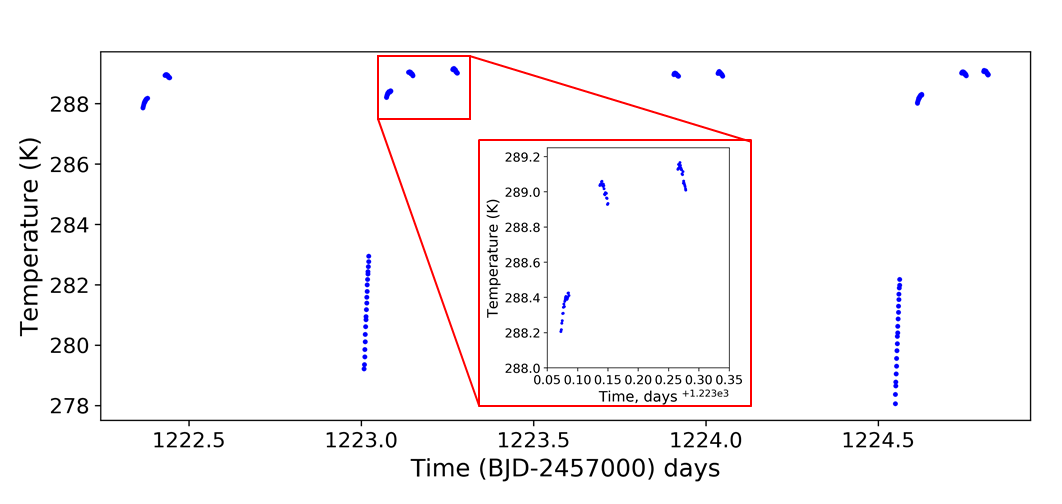}{0.45\textwidth}{Lens housing temperature}
    }\vspace{-5pt}
    \gridline{\fig{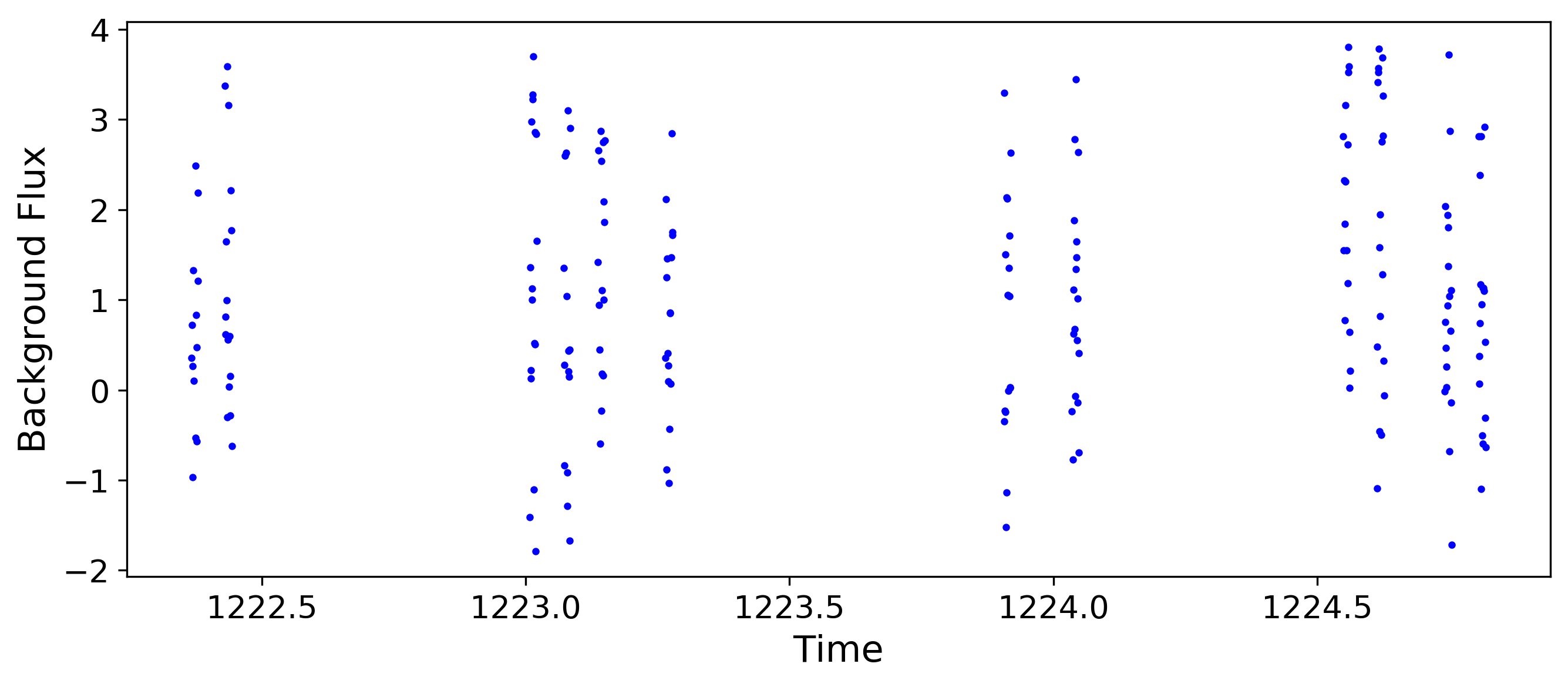}{0.45\textwidth}{Sky background}
              \fig{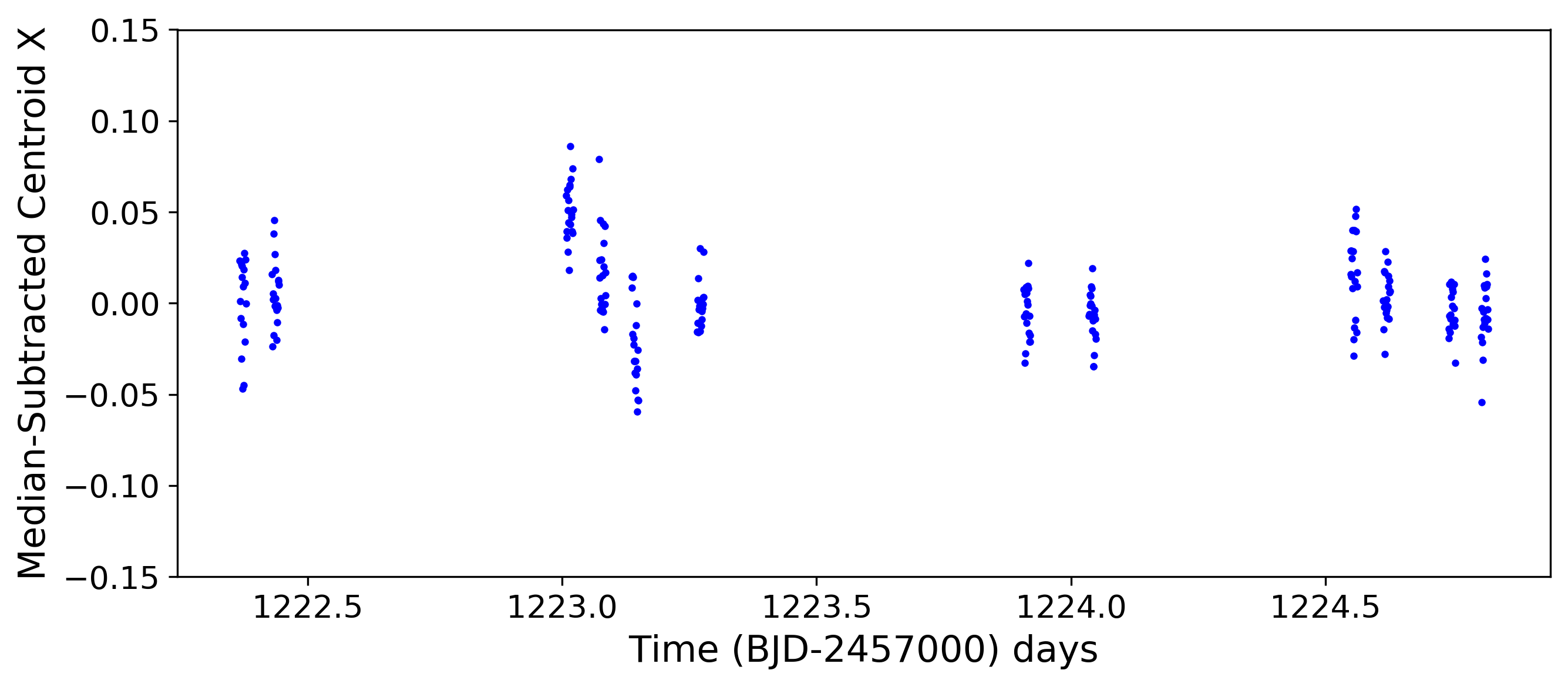}{0.45\textwidth}{Centroid X position}
    }\vspace{-5pt}
    \gridline{\fig{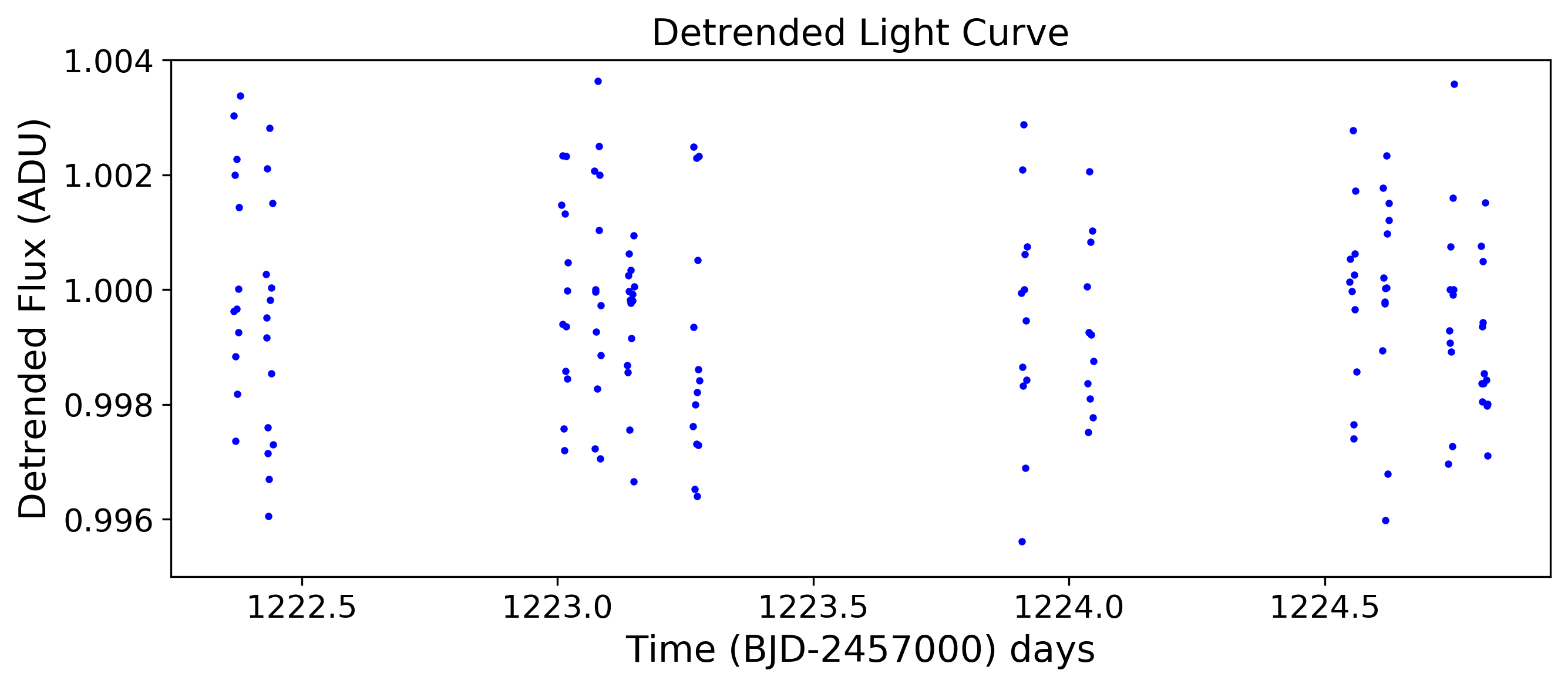}{0.45\textwidth}{Detrended lightcurve}
              \fig{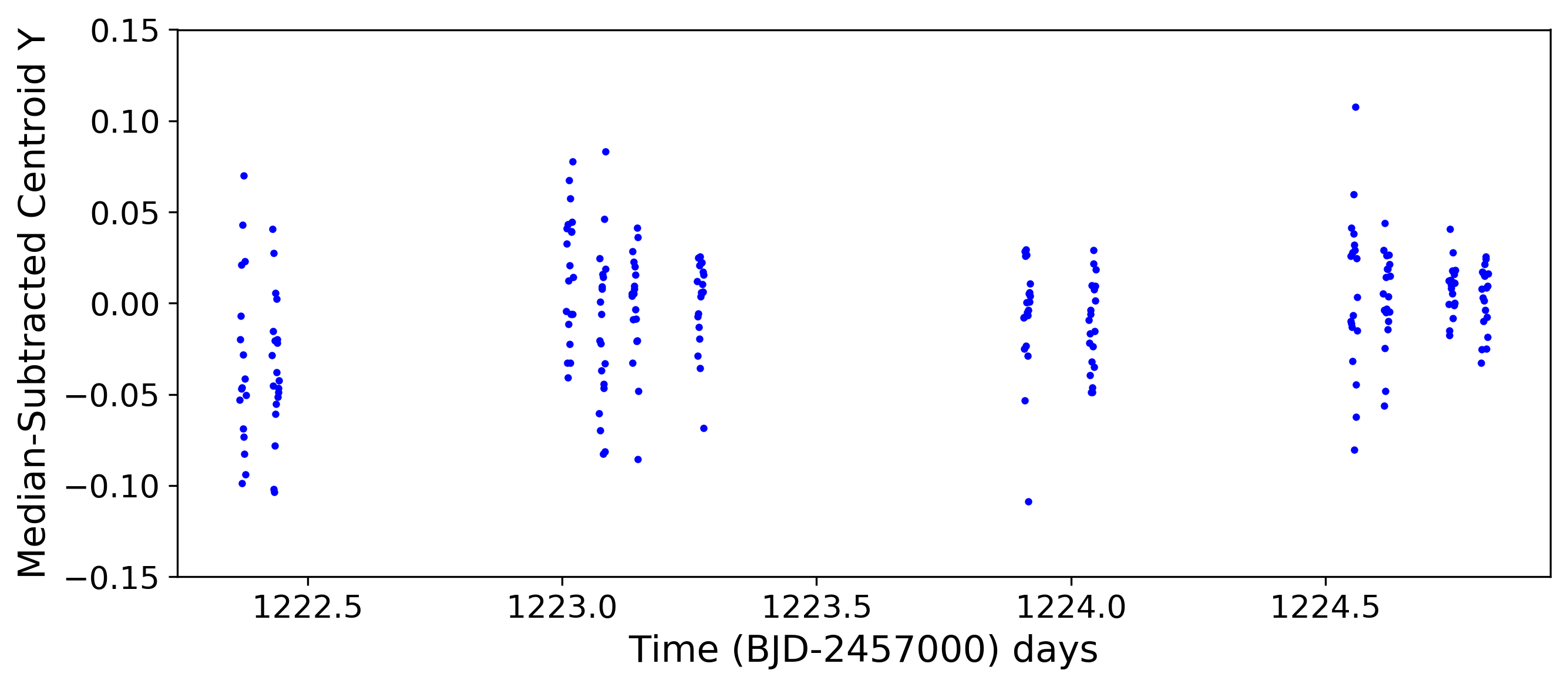}{0.45\textwidth}{Centroid Y position}
    }\vspace{-0pt}
    \caption{Initial (top left) and detrended (bottom left) lightcurves for one observation (so19).  Data used for detrending, including sky background (center left), lens housing temperature (top right), and star centroid position (center, bottom right), are shown on the same time axis as the extracted photometry.  A zoomed in inset (red) in the lens housing telemetry (top right) shows the orbit-to-orbit thermal behavior of the lens housing.  The lens housing is not under active thermal control like the detector.  The sharp rise in temperature seen in the second and fourth cluster of orbits happens when the camera first turns on and heats the lens housing; the sharp rise in temperature is not seen in the first and third sets of orbits because the camera failed to collect data during the first orbit of each set.}
    \label{fig:lc_detrend}
\end{figure*}

\subsection{Data analysis}

\subsubsection{Photometric modeling}

For lightcurve detrending, we use the Markov Chain Monte Carlo (MCMC) algorithm implementation already presented in the literature \citep{Gillon2012}. Inputs to the MCMC are the 55 Cnc photometric time-series and parameters described above. We explore different functional forms of the baseline model for each light curve (See Table~\ref{tab:results} for the results). These models can include a linear/quadratic trend with time, a second-order logarithmic ramp \citep{Knutson2007,Demory2011} usually included for telescope settling with Spitzer and HST, a polynomial of the 1) centroid position and 2) onboard temperature measured at the lens housing, as well as linear combinations of the PSF FWHM along the x and y axes. We employ the Bayesian Information Criterion \citep[BIC,][]{Schwarz1978} to discriminate between the different baseline models. In the MCMC framework, the baseline model coefficients are determined at each step using a singular value decomposition method \citep{Press1992}. During our tests, we find that the baseline model resulting in the lower Bayesian Information Criterion (Schwarz 1978) is a combination of a second order polynomial of the x/y centroid position, and a second order polynomial of the lens housing temperature. All results presented in this work have been obtained using this baseline model. We emphasise at this stage that these results are obtained over non-continuous visit durations of up to 5 hours on 55 Cnc, known to be a photometrically quiet star \citep{Fischer2007,Demory2016}. We expect the baseline model to become more complex for longer visits.

\subsubsection{Transit fit}
We fit for the baseline model described above and the 55 Cnc e transit simultaneously in a same MCMC framework. We fit for the transit centre $T_0$ and the transit depth. We include the limb-darkening linear combinations $c_1=2u_1+u_2$ and $c_2=u_1-2u_2$, where $u_1$ and $u_2$ are the quadratic coefficients drawn from the theoretical tables of \citet{Claret2011} using published stellar parameters \citep{vonBraun2011}. We impose normal priors on the  orbital period $P$, impact parameter $b$ and the limb-darkening parameters $u_1$ and $u_2$ to the values shown in Table~\ref{tab:results}. We keep the eccentricity fixed to zero \citep{Demory2012,Nelson2014}. We execute two Markov chains of 50,000 steps each and assess their efficient mixing and convergence using the Gelman-Rubin statistic \citep{Gelman1992} by ensuring $r <1.001$. Results for this MCMC fit are shown in Table~\ref{tab:results}.

\subsubsection{Noise properties}
\begin{figure*}
    \centering
    \gridline{\fig{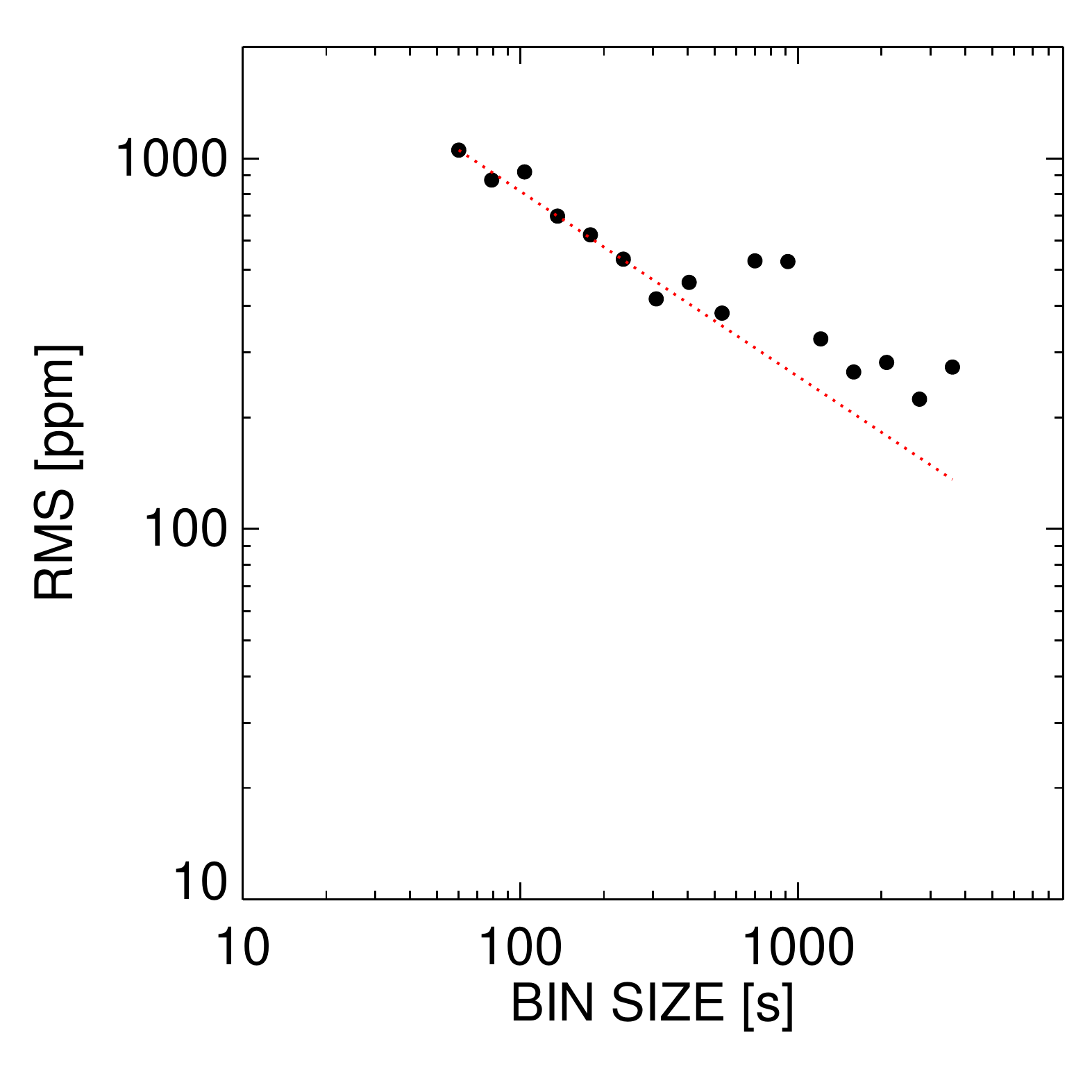}{0.2\textwidth}{td14}
              \fig{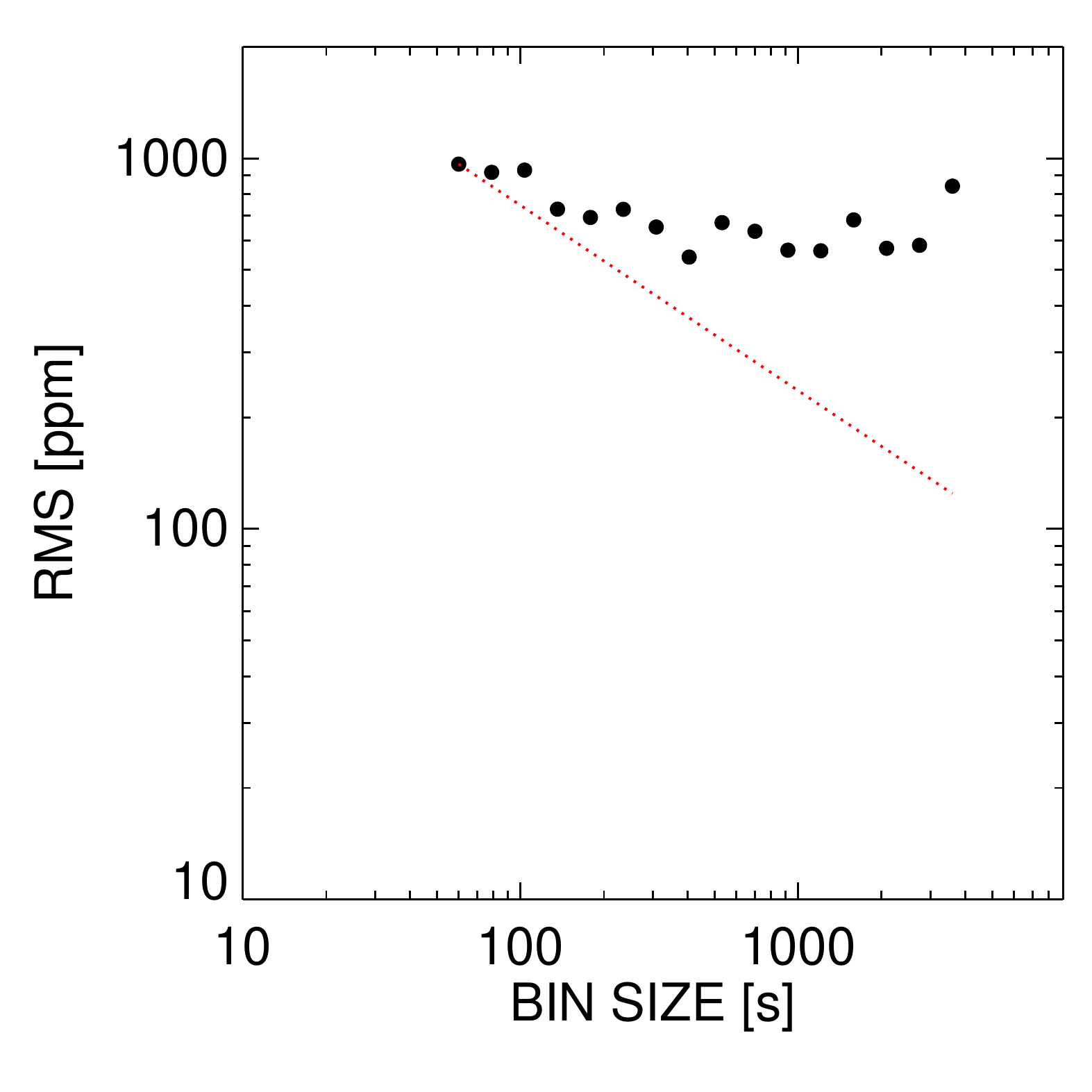}{0.2\textwidth}{td16}
              \fig{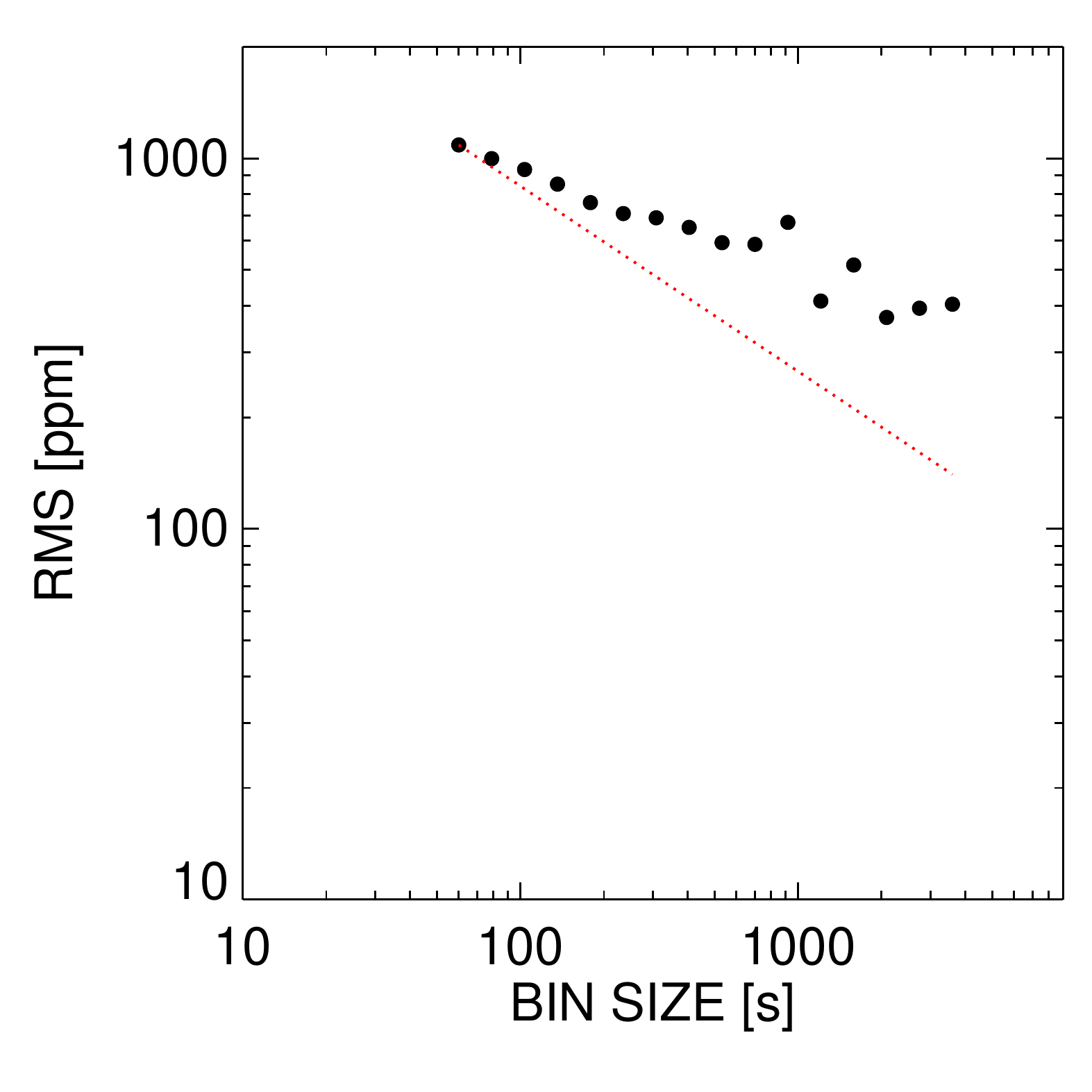}{0.2\textwidth}{td18}
              \fig{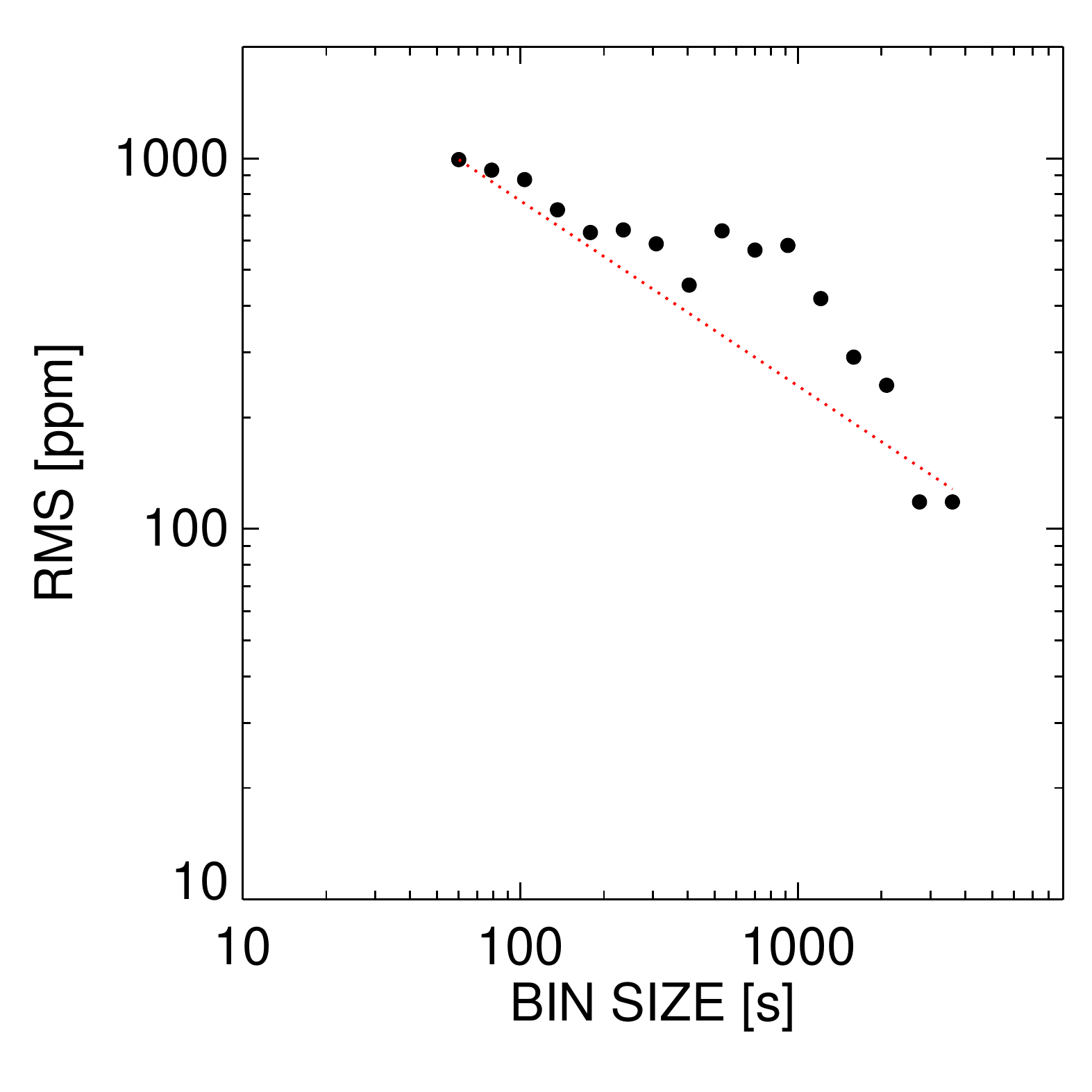}{0.2\textwidth}{td23}
    }\vspace{-5pt}
    \gridline{\fig{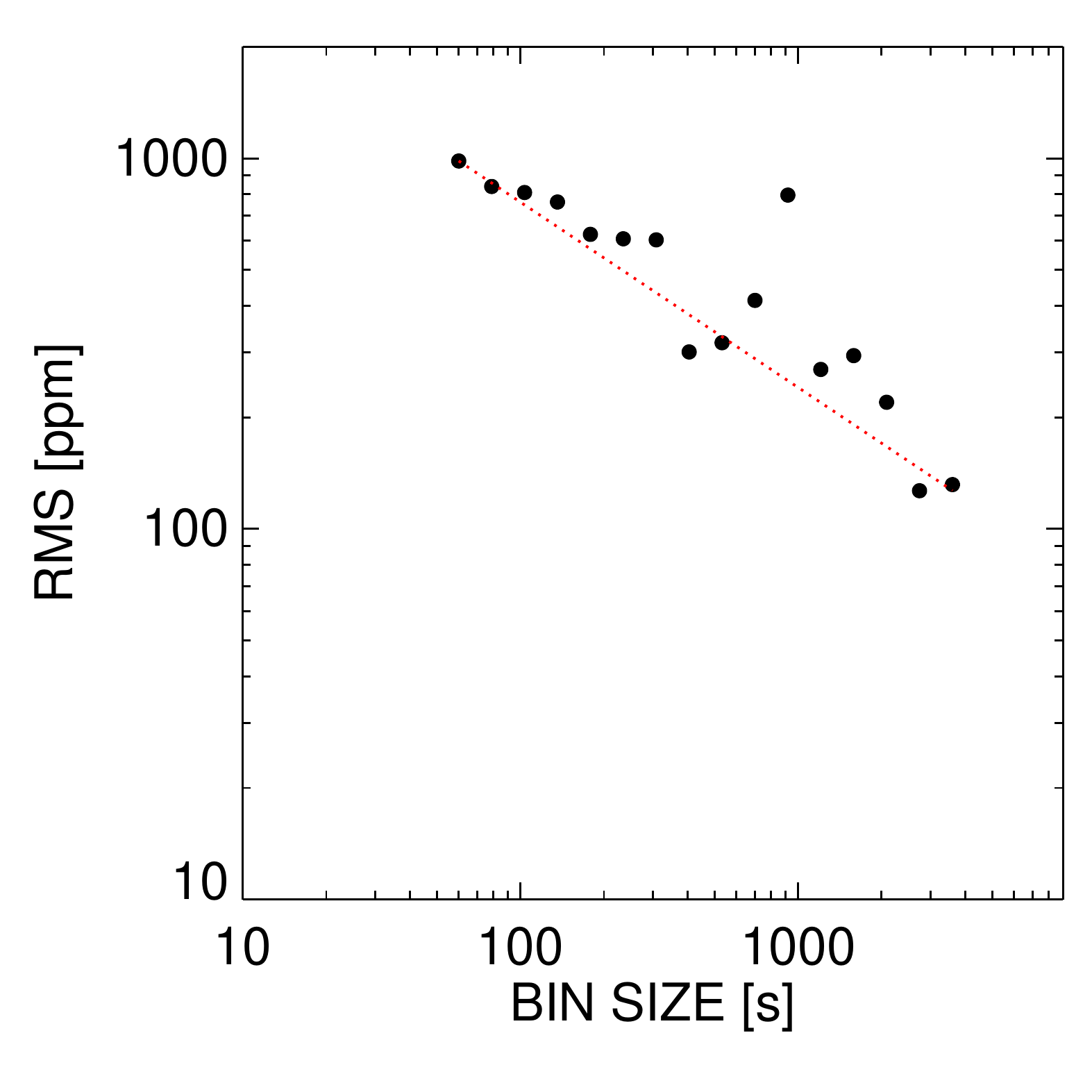}{0.2\textwidth}{td24}
              \fig{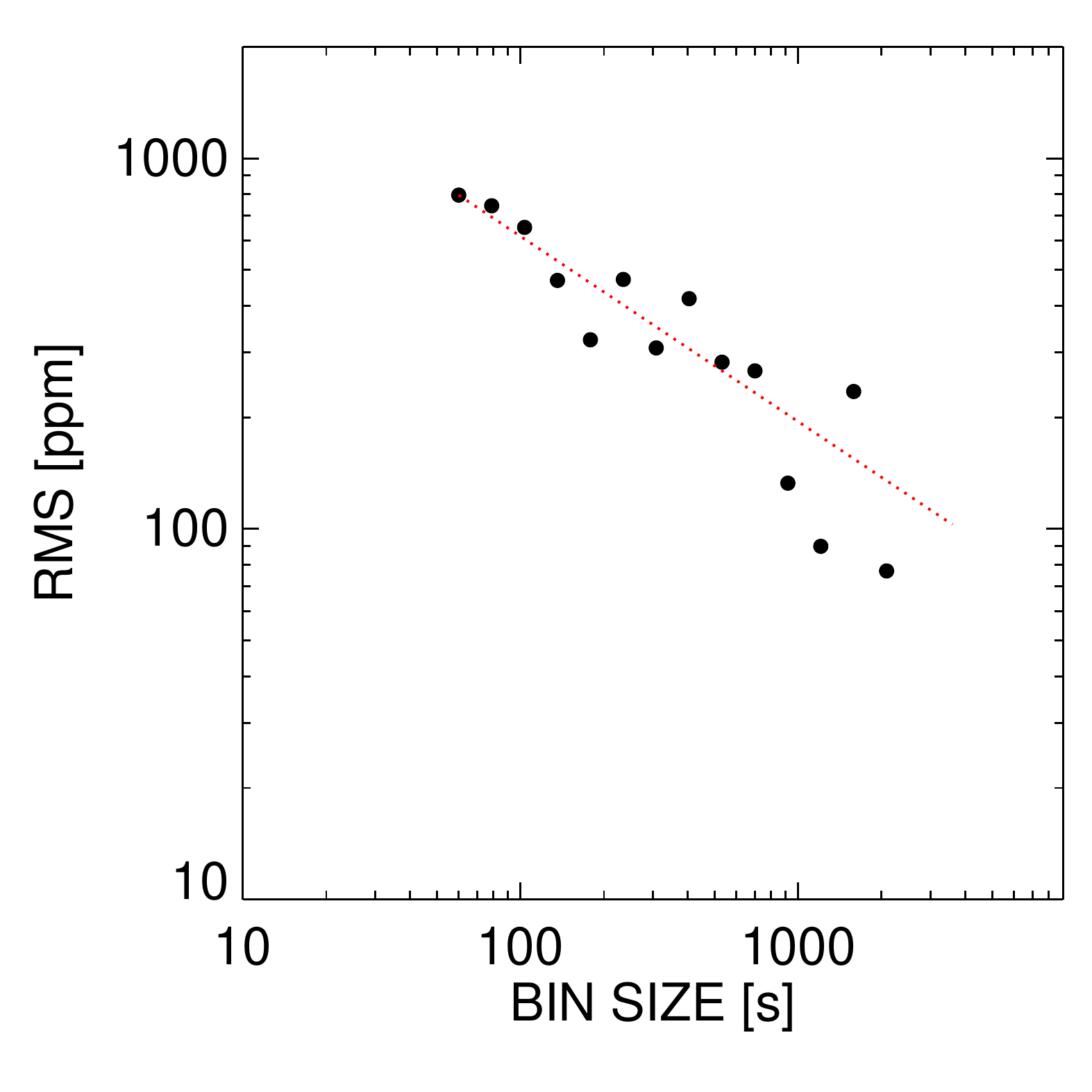}{0.2\textwidth}{so19, orbits 1-2}
              \fig{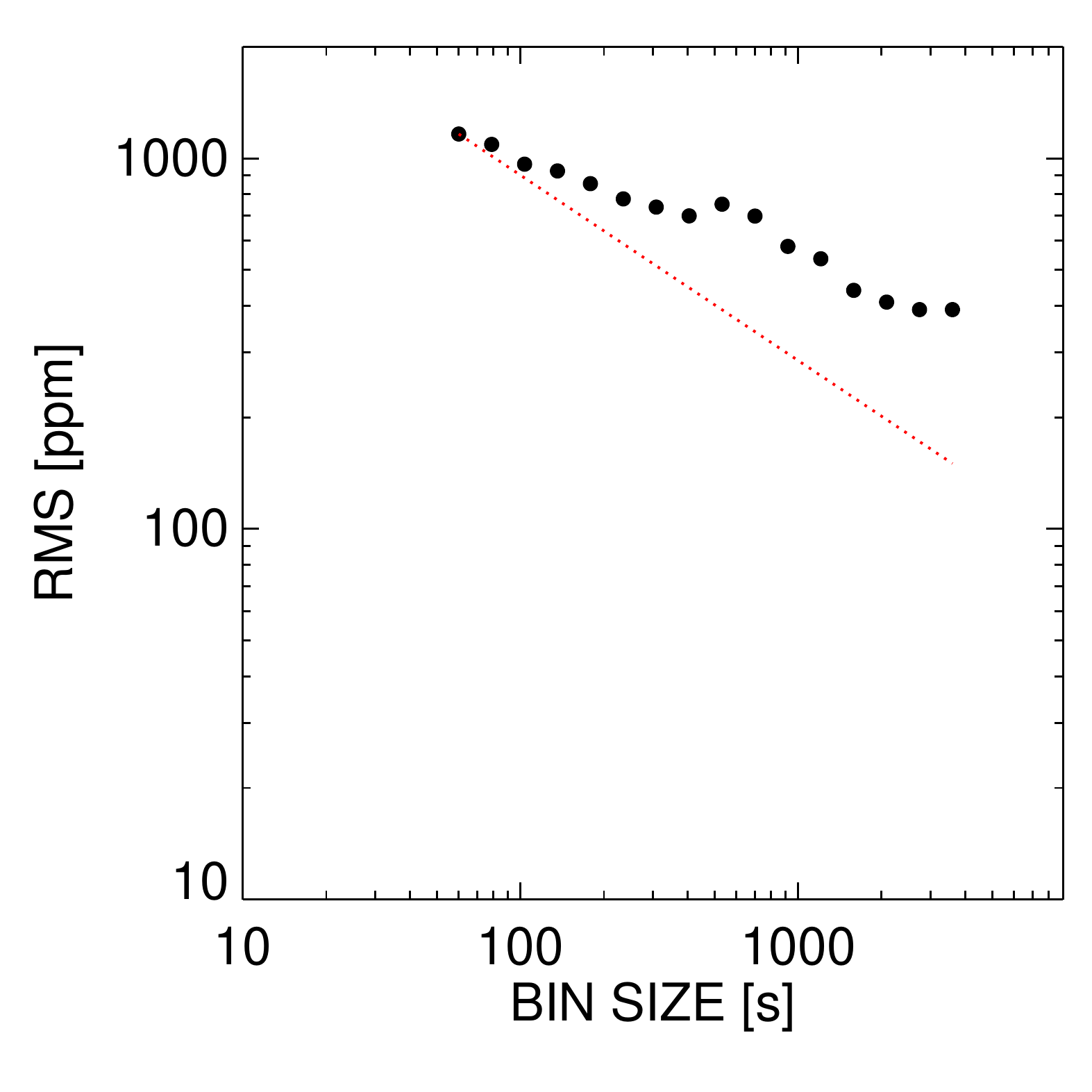}{0.2\textwidth}{so19, orbits 3-6}
              \fig{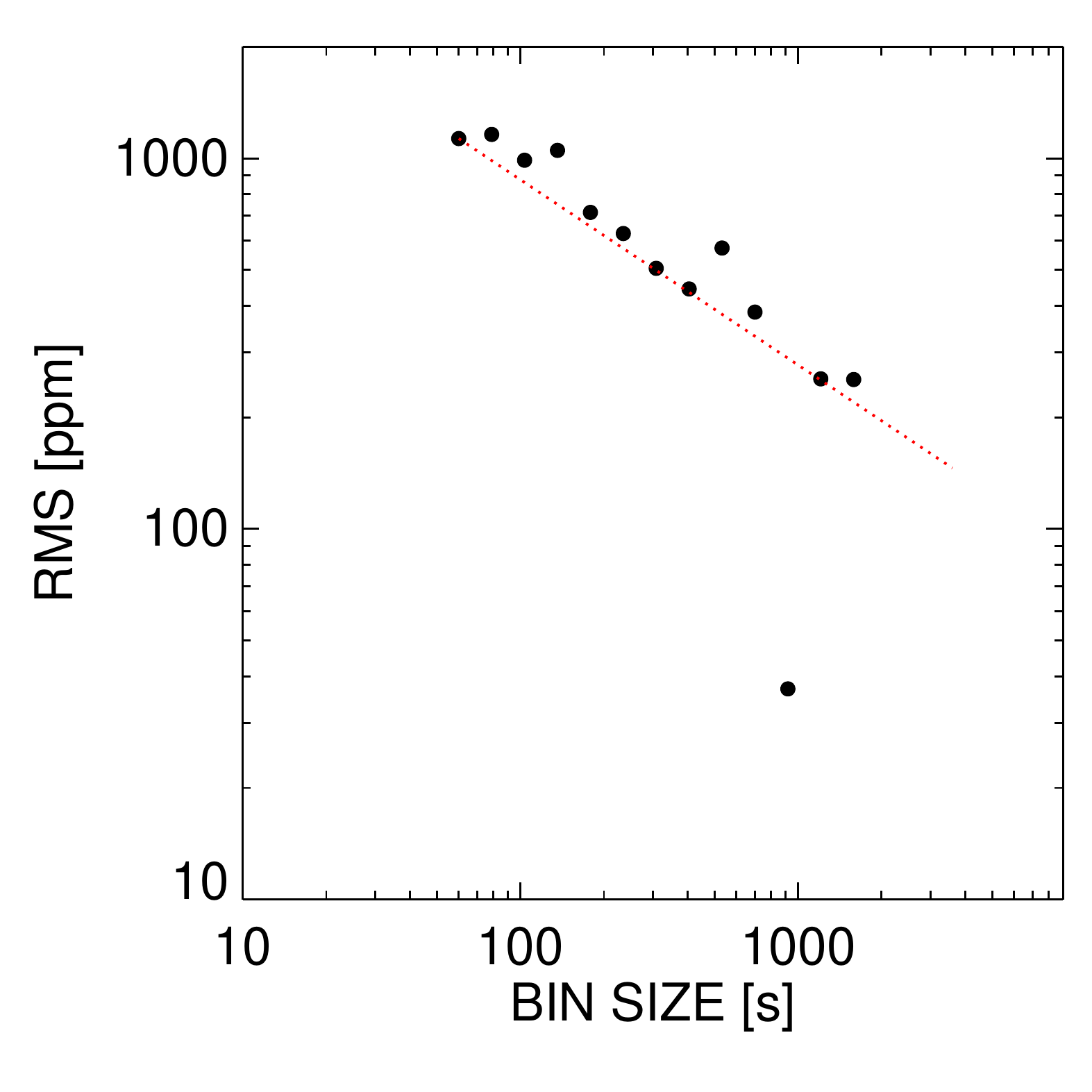}{0.2\textwidth}{so19, orbits 7-8}
    }\vspace{-5pt}
    \gridline{\fig{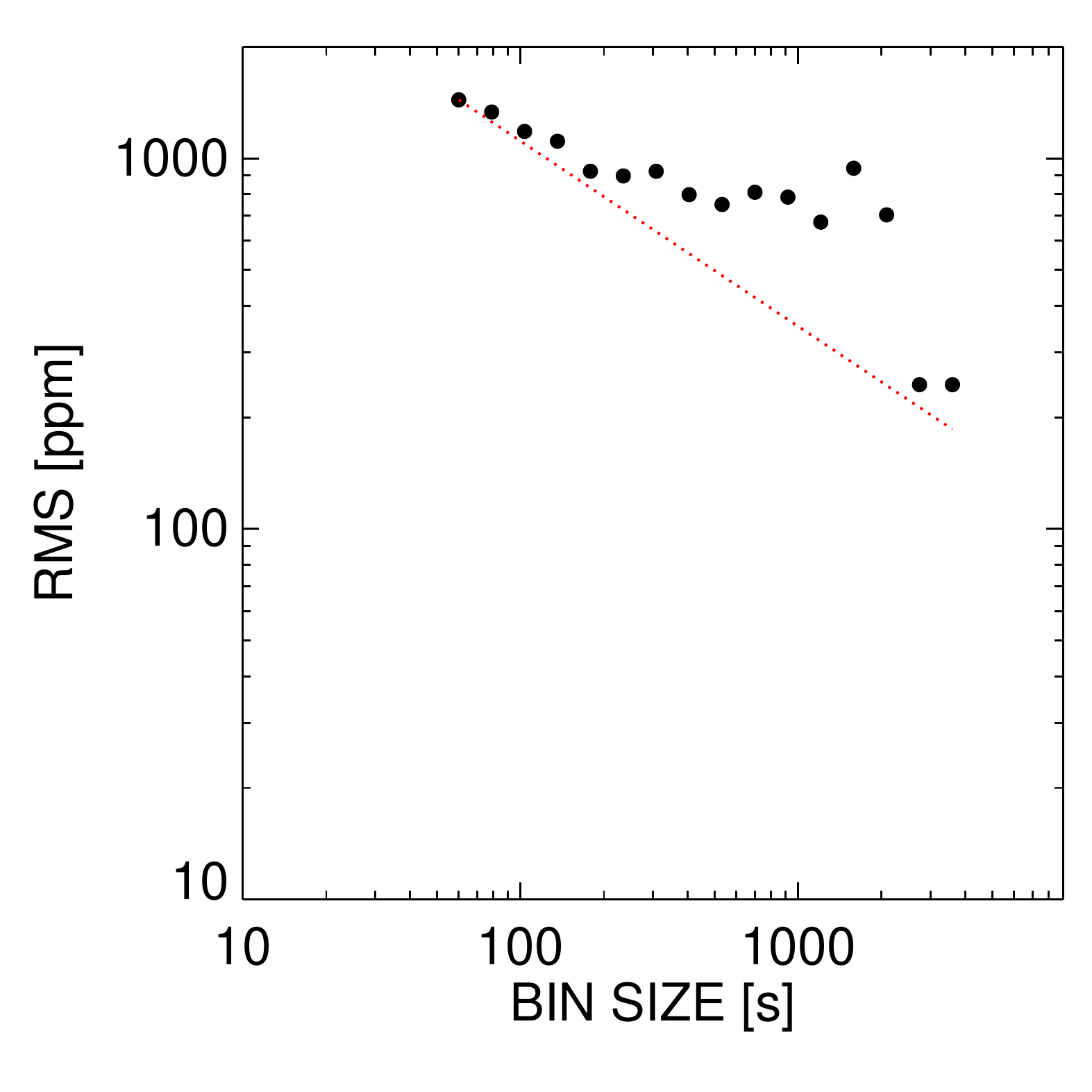}{0.2\textwidth}{so19, orbits 9-12}
              \fig{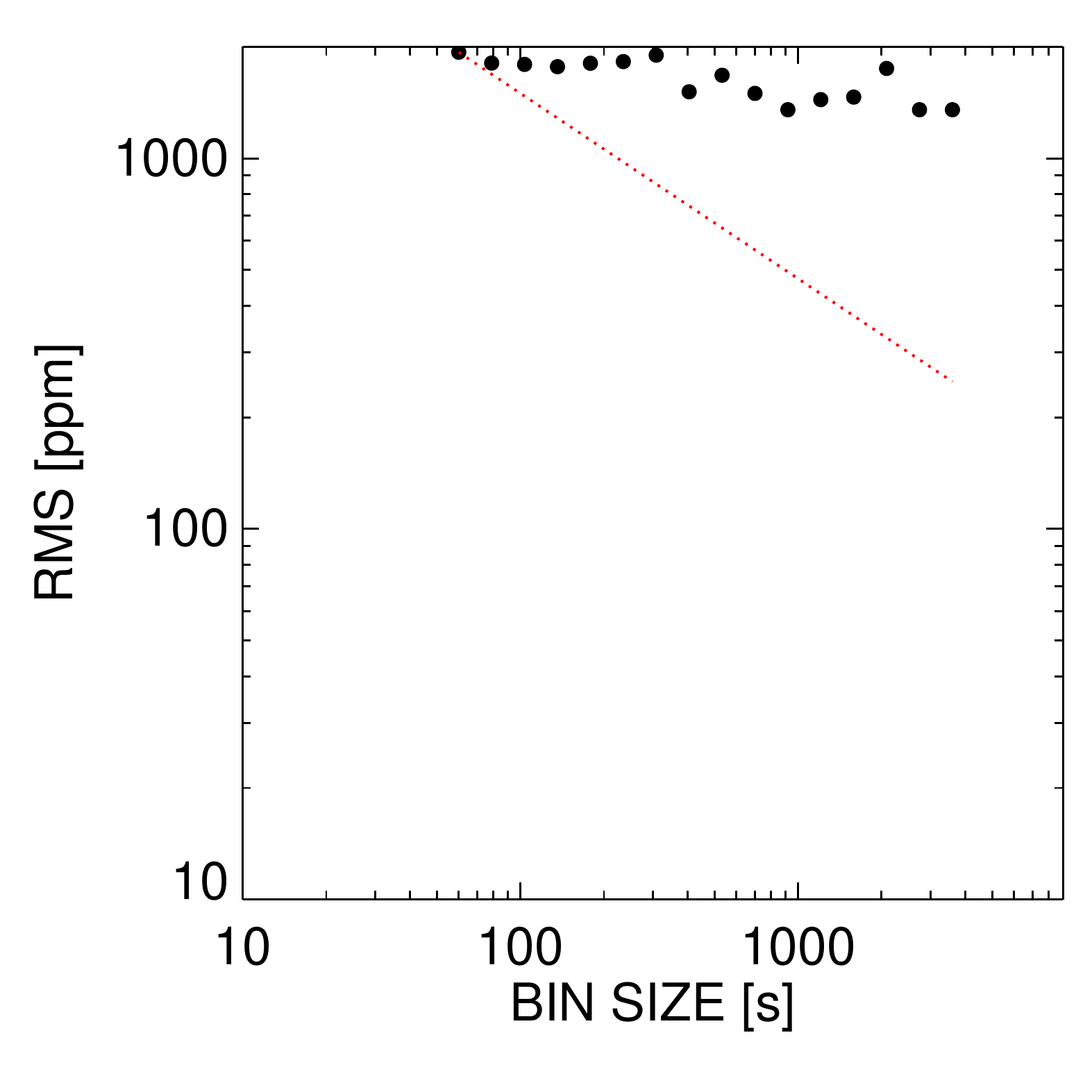}{0.2\textwidth}{so20}
              \fig{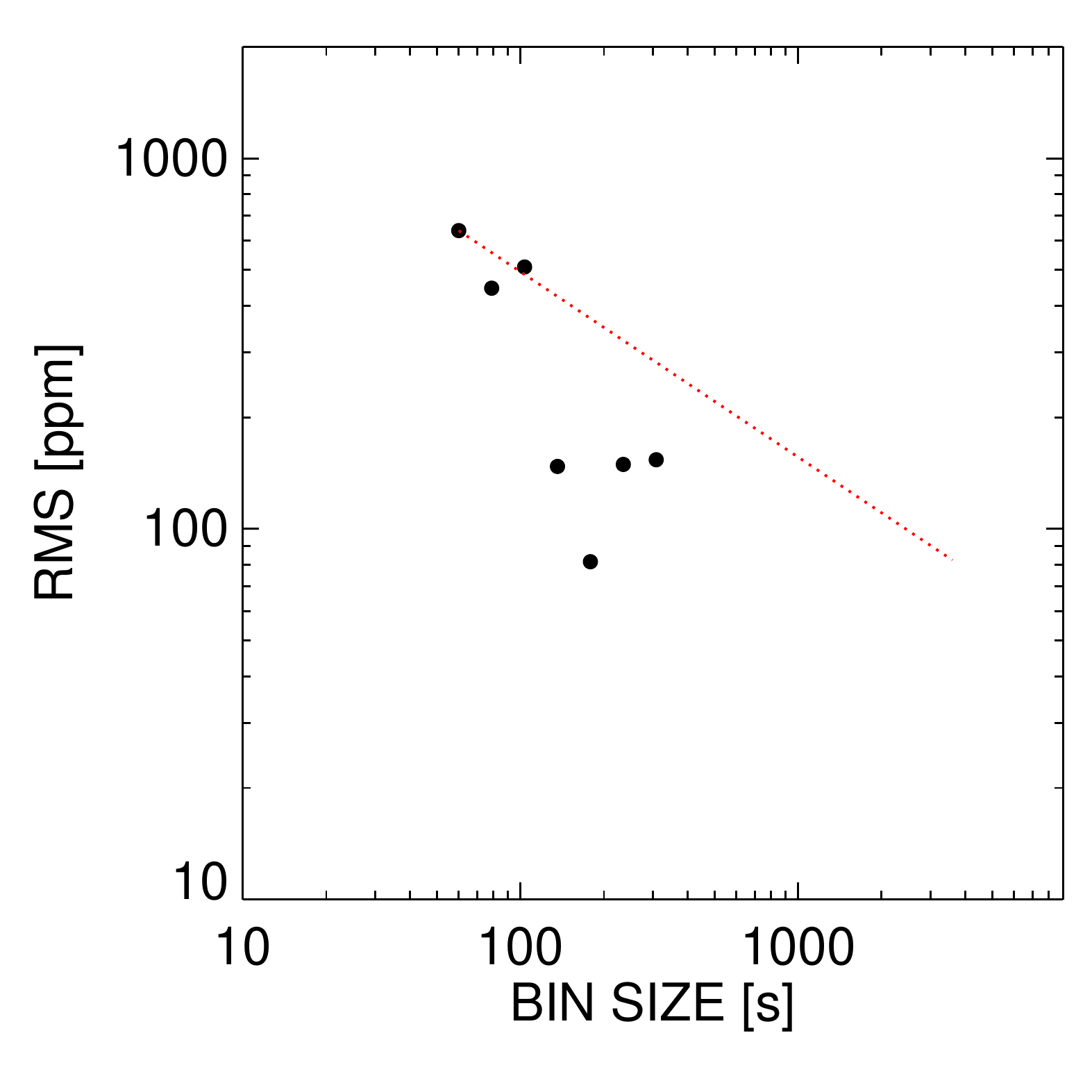}{0.2\textwidth}{so24, orbit 3}
              \fig{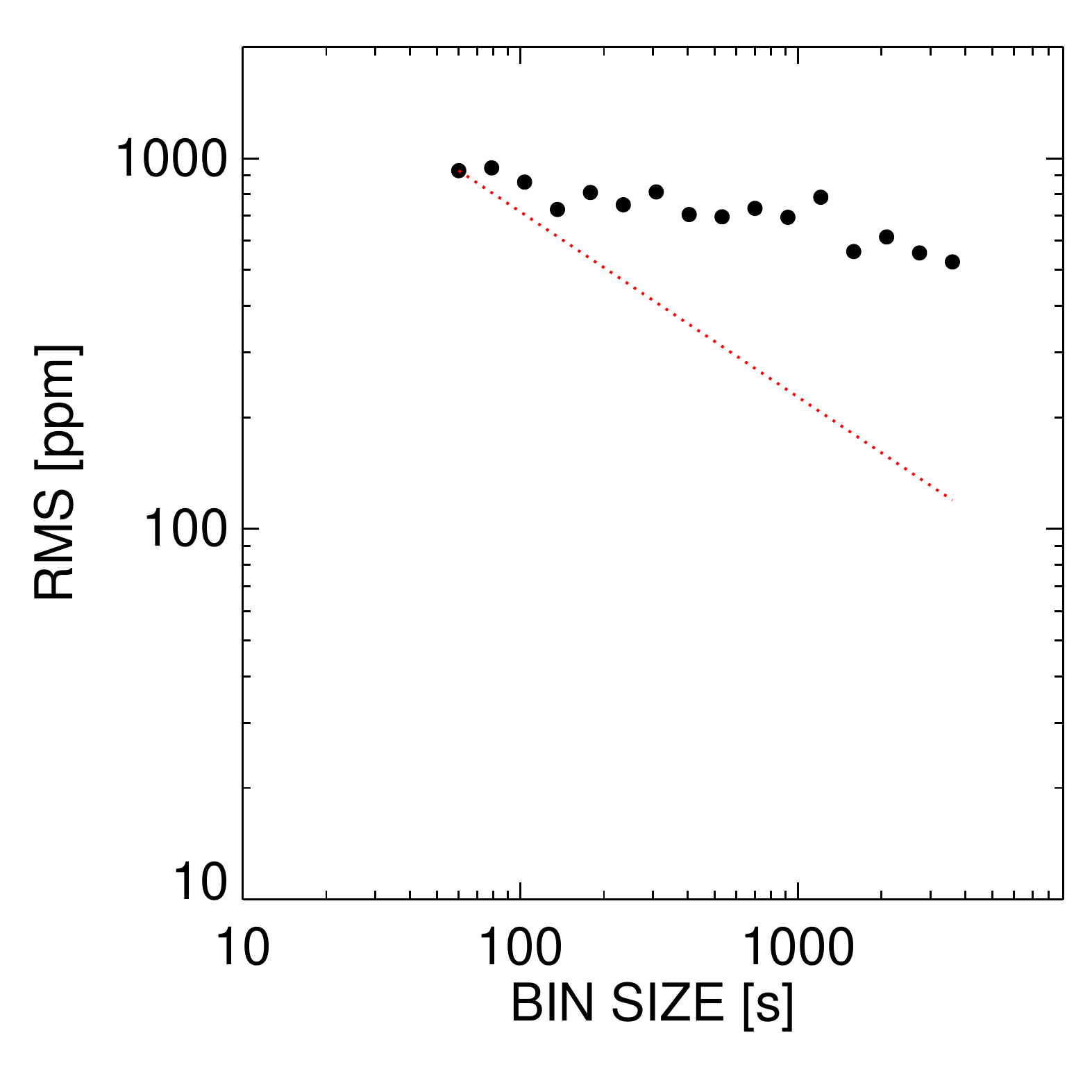}{0.2\textwidth}{so25, orbit 1-3}
    }\vspace{-5pt}
    \gridline{\fig{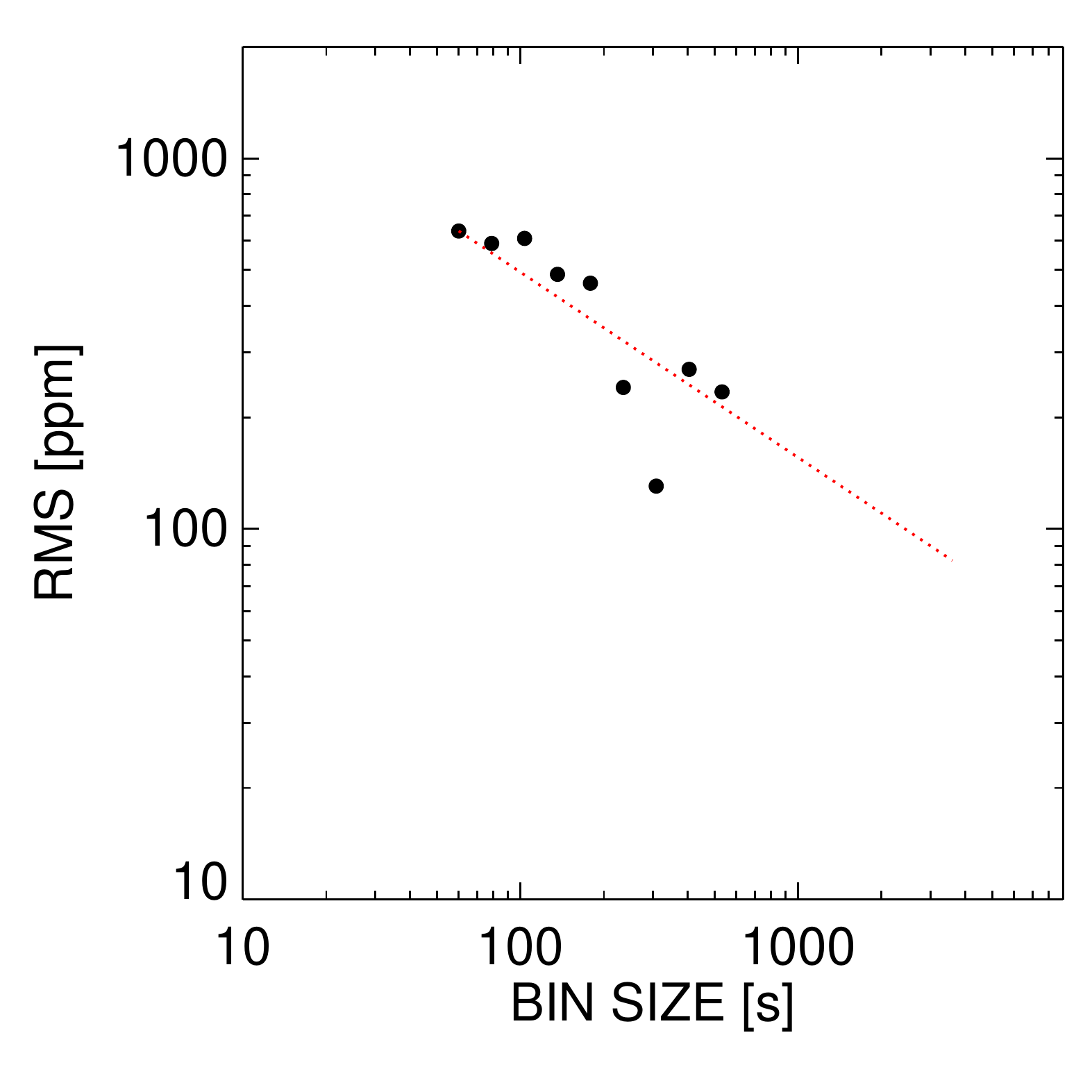}{0.2\textwidth}{so25, orbit 4}
              \fig{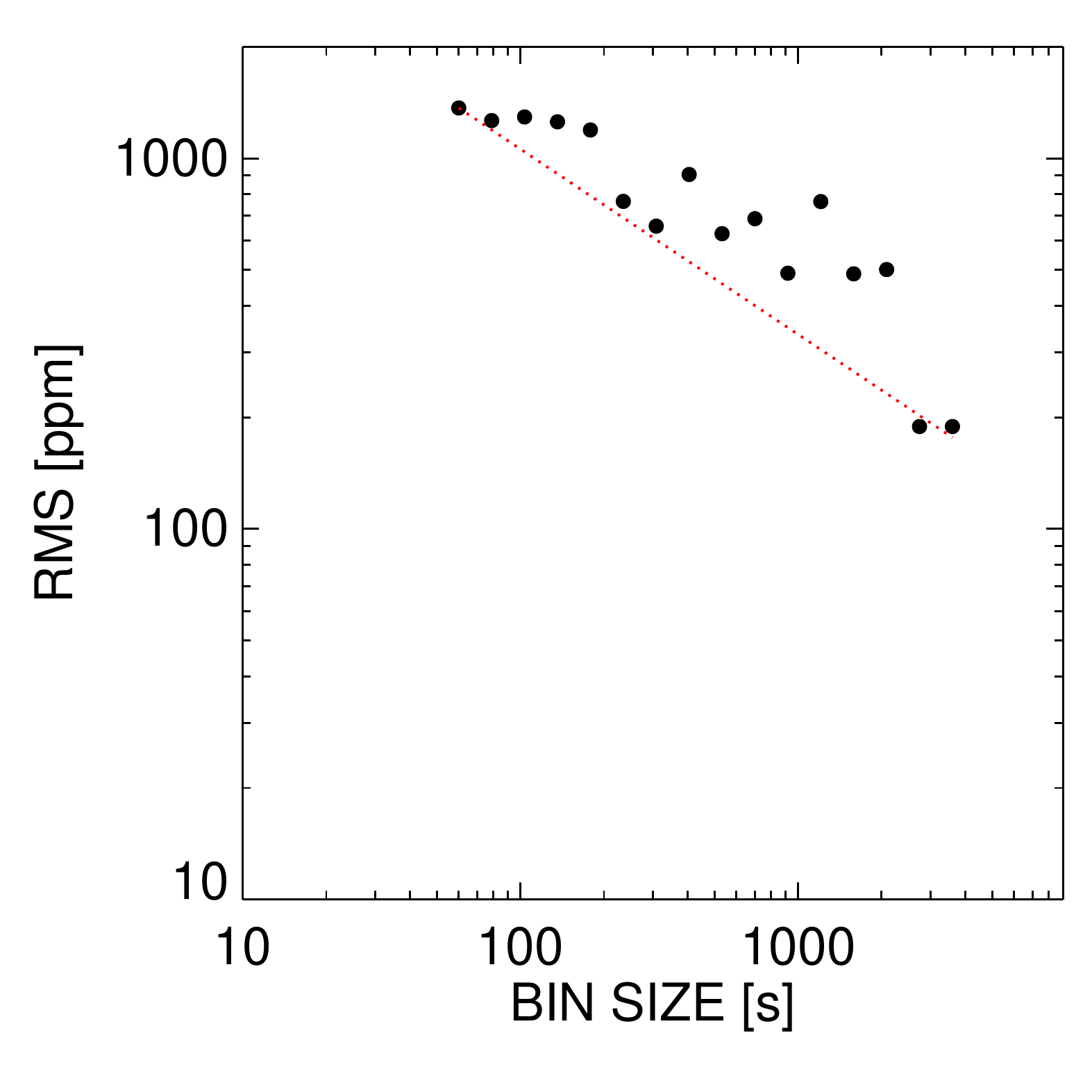}{0.2\textwidth}{so25, orbits 5-7}
              \fig{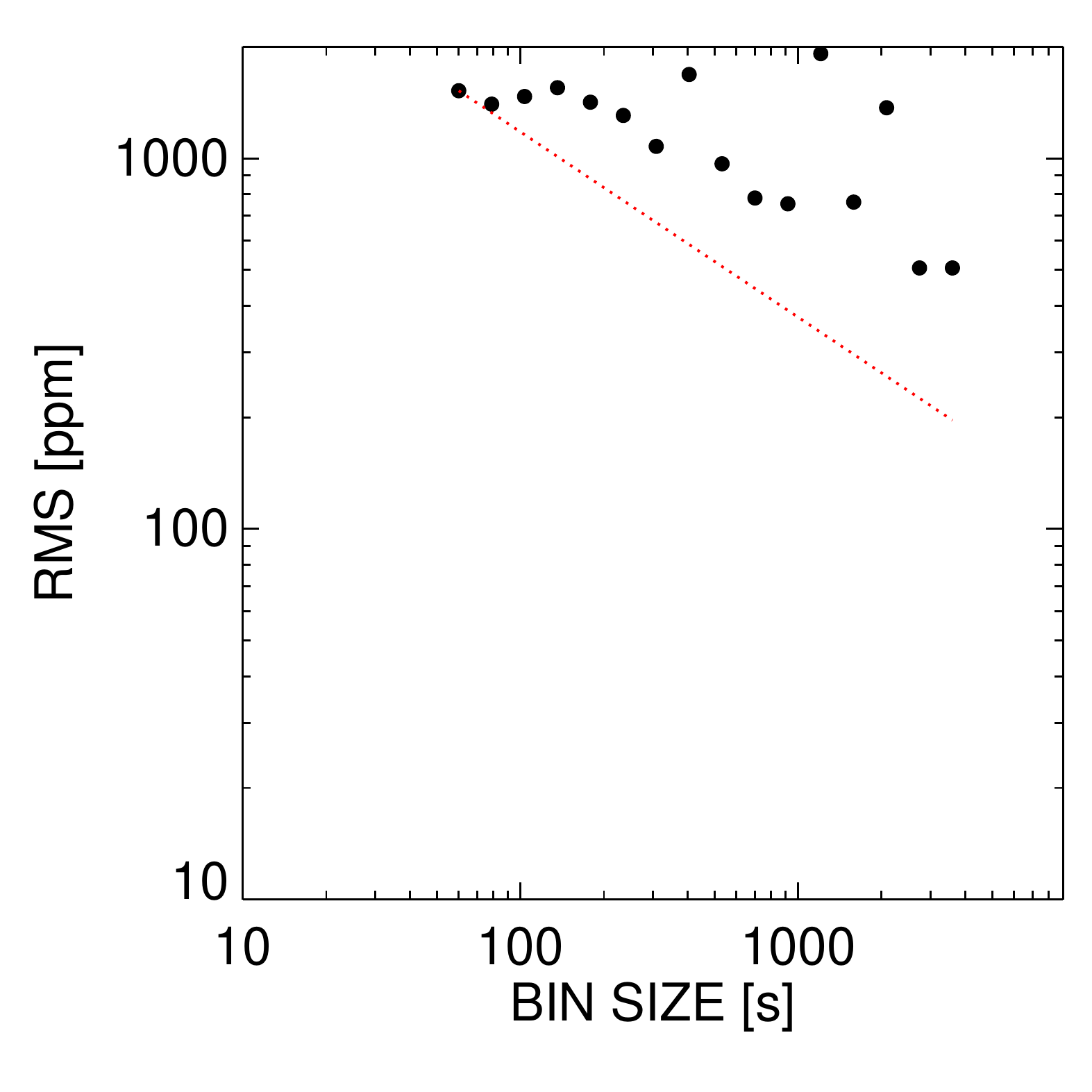}{0.2\textwidth}{so26}
              \fig{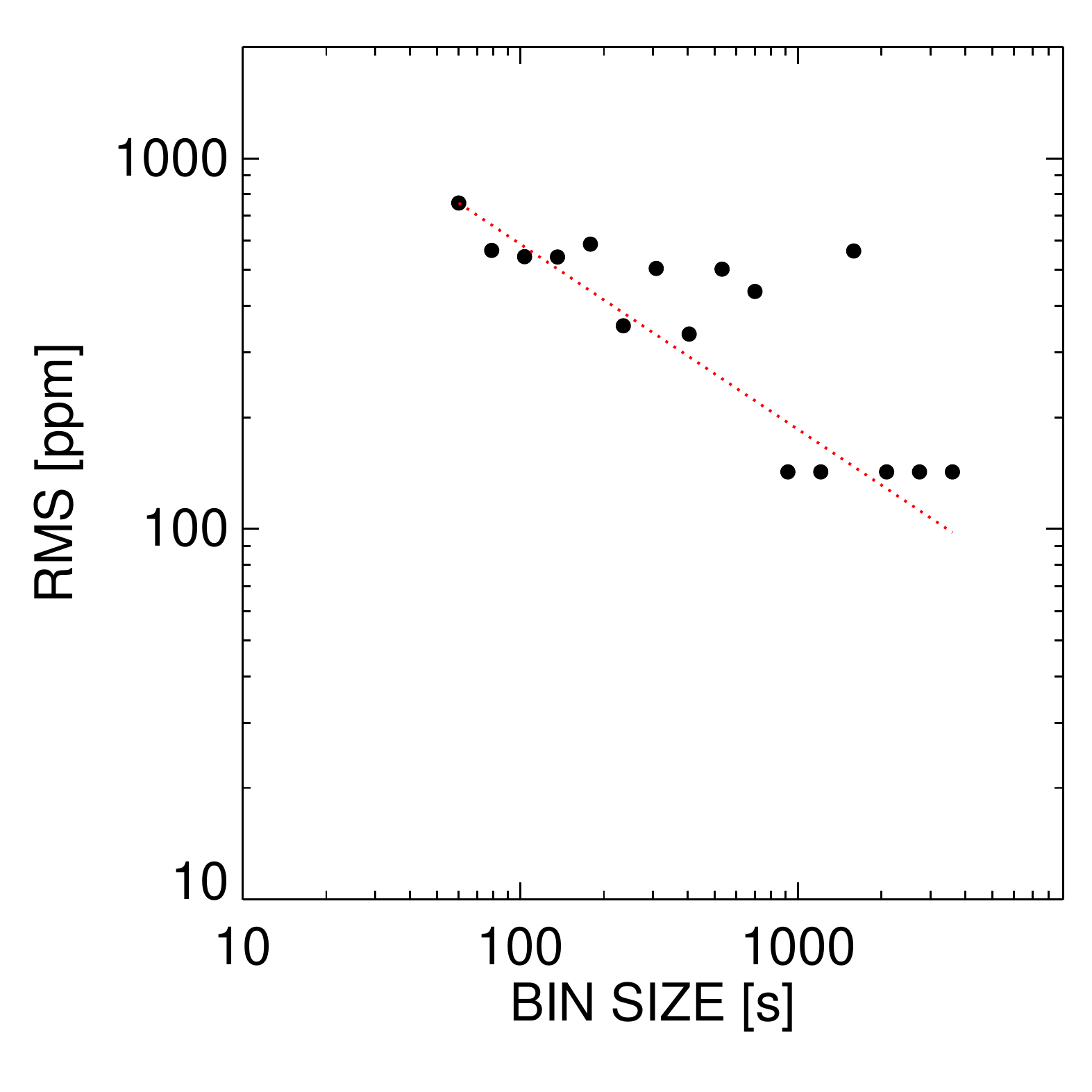}{0.2\textwidth}{so28, orbits 6-8}
    }\vspace{-5pt}
    \gridline{\fig{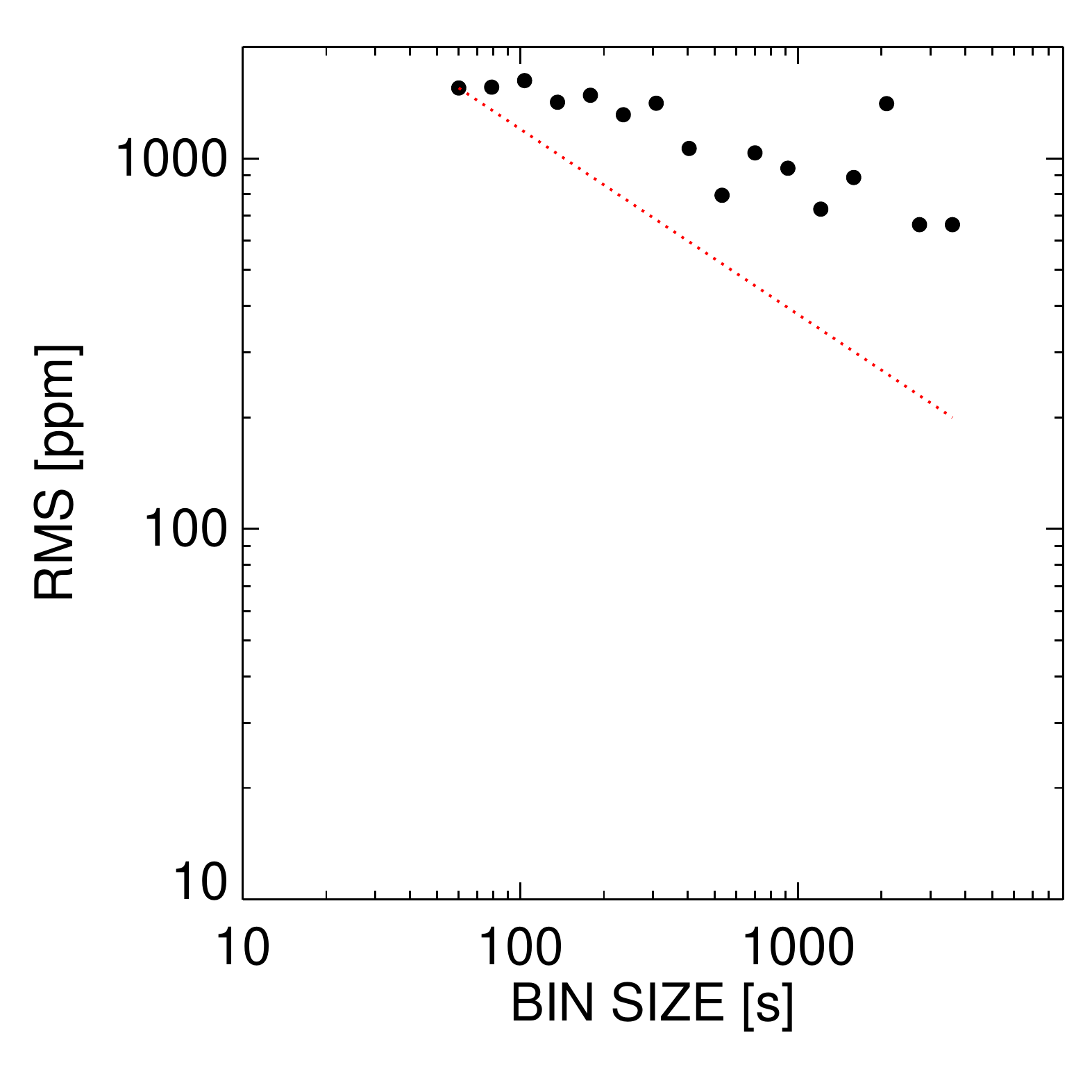}{0.2\textwidth}{so30, orbits 1-4}
              \fig{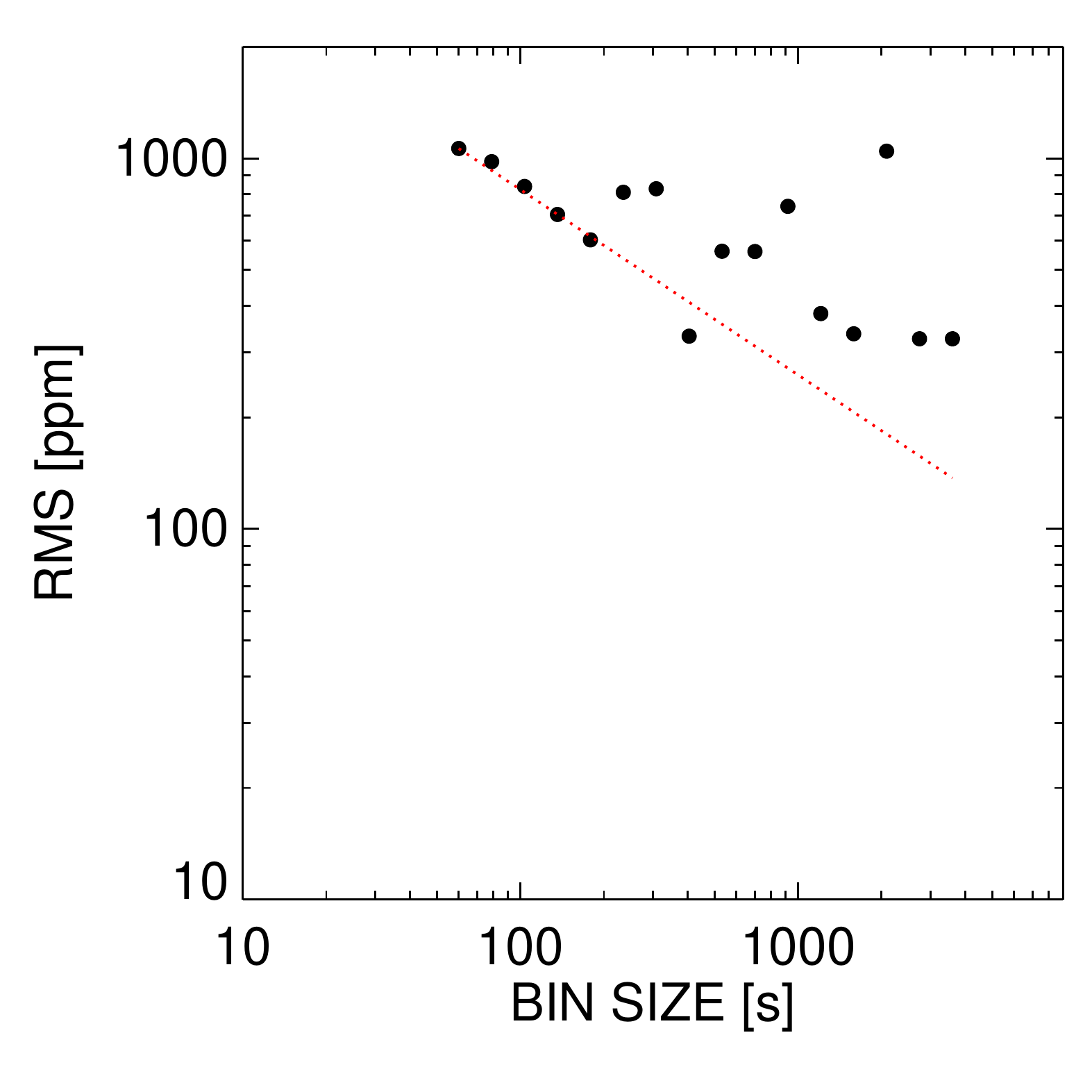}{0.2\textwidth}{so30, orbits 5-8}
    }
    \caption{Photometric RMS versus bin size for all data sets. Black filled circles indicate the photometric residual RMS for different time bins. Each panel corresponds to an individual data set (see Table~\ref{tab:listobs}). The expected decrease in Poisson noise normalized to an individual bin (1 min) precision is shown as a red dotted line. }
    \label{fig:rms}
\end{figure*}

The detector's average gain value of 6.44~e-/ADU, combined to the ADU measured counts translate to a photon-noise limit of about 400~ppm/min for 55 Cnc, depending on the visit. Our observations yield residual RMS between 620~ppm/min  and 1450~ppm/min. The analysis of this noise excess shows that correlated noise contributes an average of 42\% to the photometric noise budget over 15 to 100 min timescales. We attribute the remaining noise to astrophysical/instrumental noise that has not been characterized at this stage. We show in Figure~\ref{fig:rms} the behavior of the photometric RMS with bin size, demonstrating the nominal contribution from correlated noise in the data.

\section{55 Cancri e transit search} \label{sec:results}
The 55 Cancri e transit was chosen as a demonstration of ASTERIA's photometric capabilities since the transit ephemeris is known.  We analyzed the data first using the literature values for the planet impact parameter ($b$), orbital period ($P$), and transit $T_0$ as priors, and then removed the priors to determine the transit detection efficiency in a blind search.

\subsection{Transit search} \label{sec:priors}
Our photometric analysis of the 55 Cnc ASTERIA data yields a transit detection at the $\sim$2.2$\sigma$ level. We conducted two MCMC fits: with and without priors on $T_0$. Both fits used the priors on $b$ and $P$ listed in Table \ref{tab:results}. When using the published $T_0$  as a prior \citep{Sulis2019}, we find a transit depth median value of 374$\pm$170 ppm (Figure \ref{fig:full_lc}, top). We show the probability distribution function of the retrieved transit depth in Figure~\ref{fig:pdf}. This is in agreement with the values published in the literature at similar wavelengths: \citet{Winn2011} found 380$\pm$52 ppm with MOST, \citet{DeMooij2014} 361$\pm$49 ppm between 457 and 671nm, 331$\pm$36 ppm with HST STIS/G750L \citep{Bourrier2018} and 346$\pm$15 ppm in the global anaylsis of MOST data \citep{Sulis2019}. Table \ref{tab:results} lists all the retrieved parameters for the fit with a prior on $T_0$.  Without a prior on $T_0$ , we recover a transit depth median value of 637$\pm$235 ppm and a transit center $T_0 = 8200.390^{+0.01}_{-0.01}$ (Figure~\ref{fig:full_lc} (center)). The deviations in these two parameters are allowed, because the fit prefers parameters such that there is no coverage of the ingress and egress phases of the transit, leaving the transit parameters less constrained. The resulting best-fit value for the $T_0$ is off by 63~min compared to the most recent published ephemeris of 55 Cnc e \citep{Sulis2019}. 

\subsection{Blind transit search and injection-recovery test} \label{sec:blind}
\begin{figure*}[h]
\centering
\includegraphics[width=0.75\textwidth]{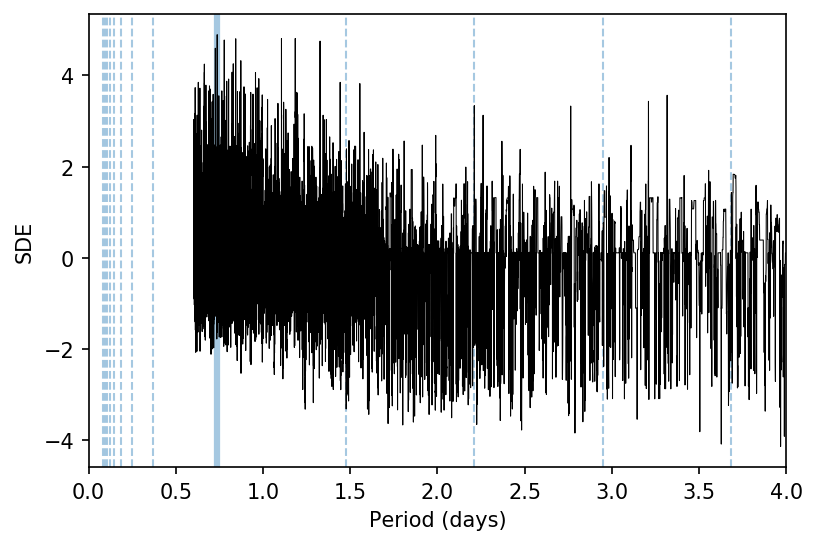}
\caption{TLS Periodogram of the ASTERIA 55 Cnc e timeseries. The peak in the spectrum is marked in blue and matches 55 Cnc e's orbital period. The location of possible aliases are indicated in dash blue vertical lines.}
\label{fig:periodogram}
\end{figure*}

\begin{figure*}[h!]
    \centering
    \includegraphics[width=0.6\textwidth]{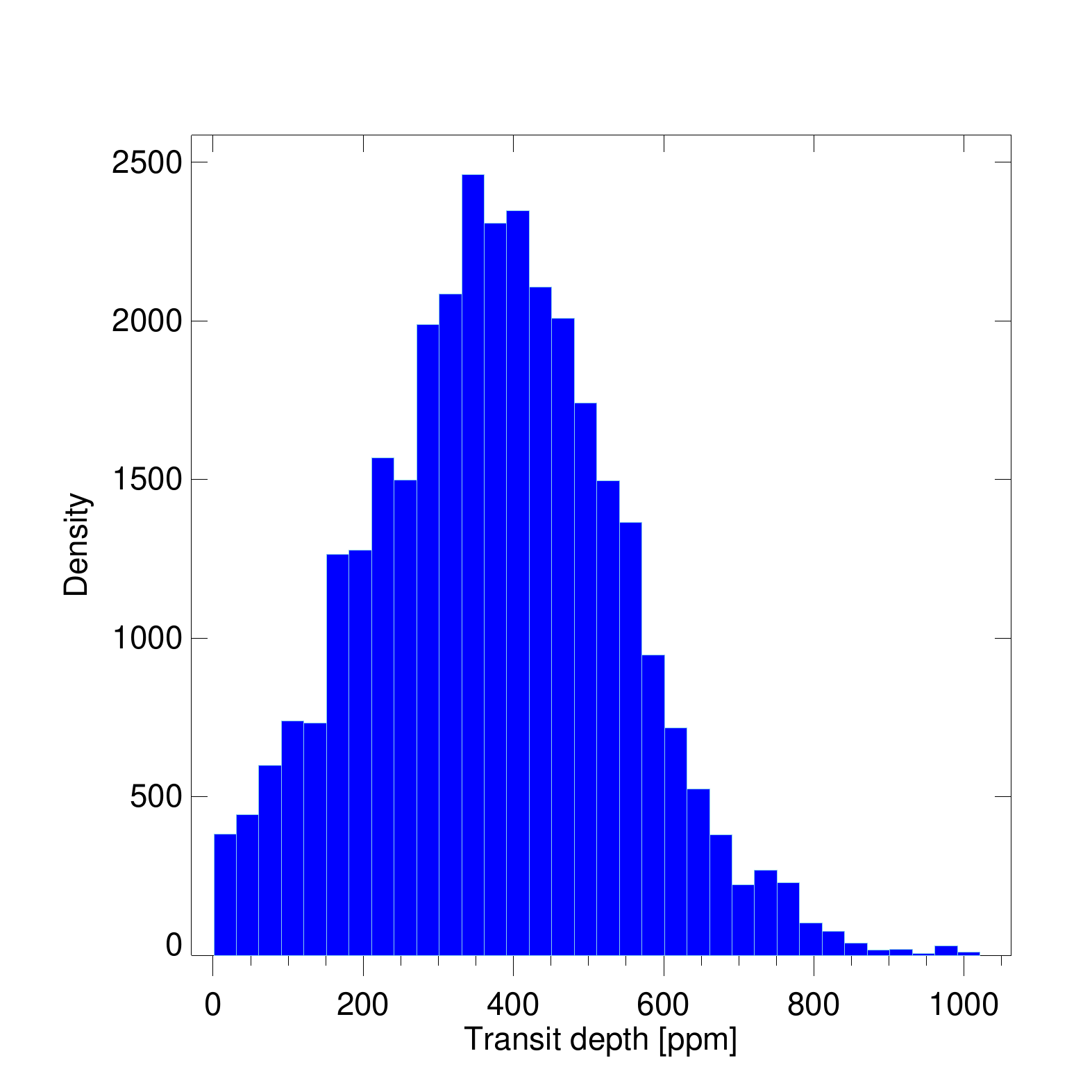}
    \caption{Posterior distribution function of the transit depth of 55 Cnc e recovered from ASTERIA photometric observations, given priors on $T_0$, $b$, and $P$.}
    \label{fig:pdf}
\end{figure*}

\begin{figure*}[h]
\gridline{\fig{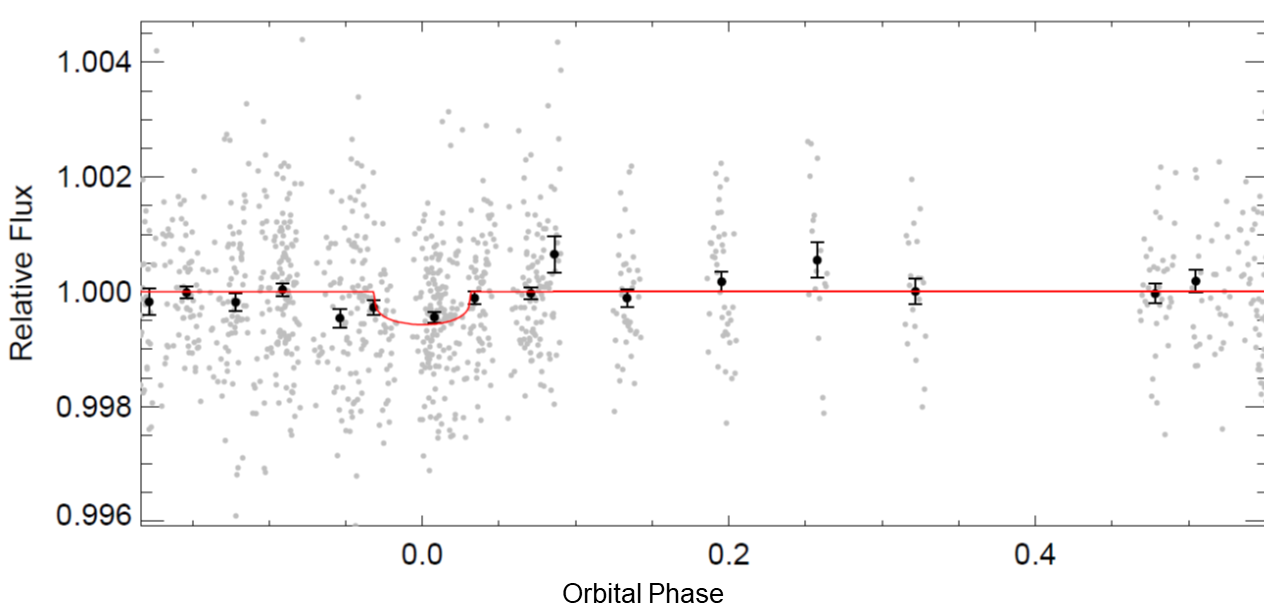}{0.75\textwidth}{a}
}\vspace{-10pt}
\gridline{\fig{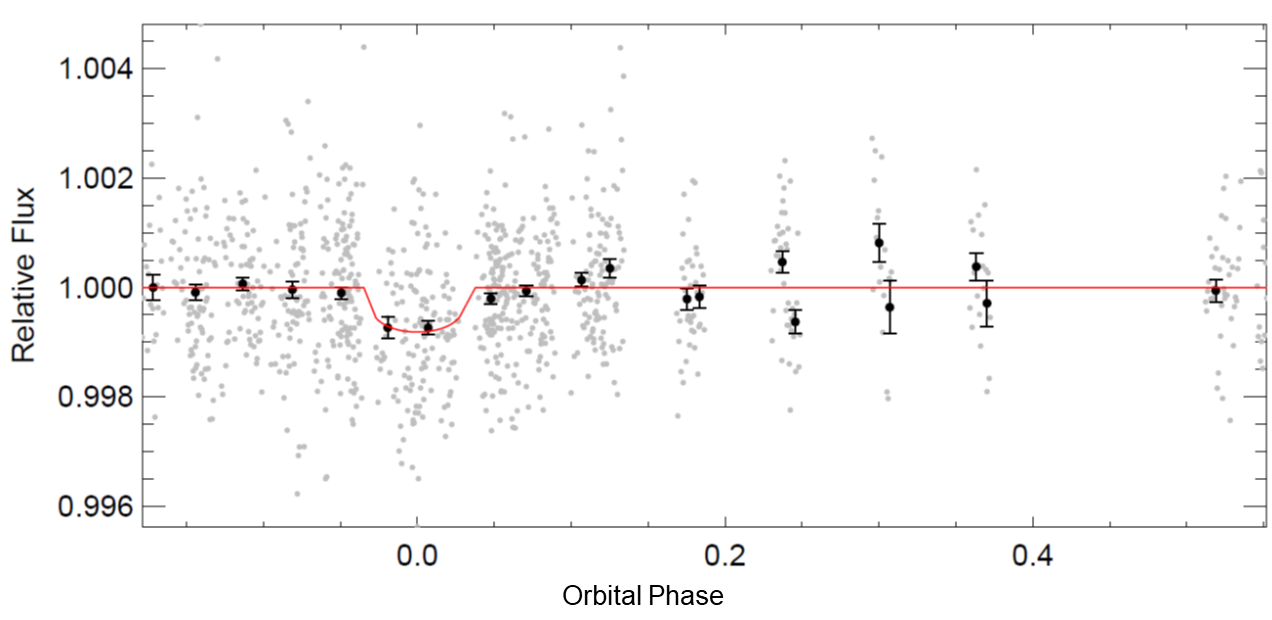}{0.75\textwidth}{b}
}\vspace{-10pt}
\gridline{\fig{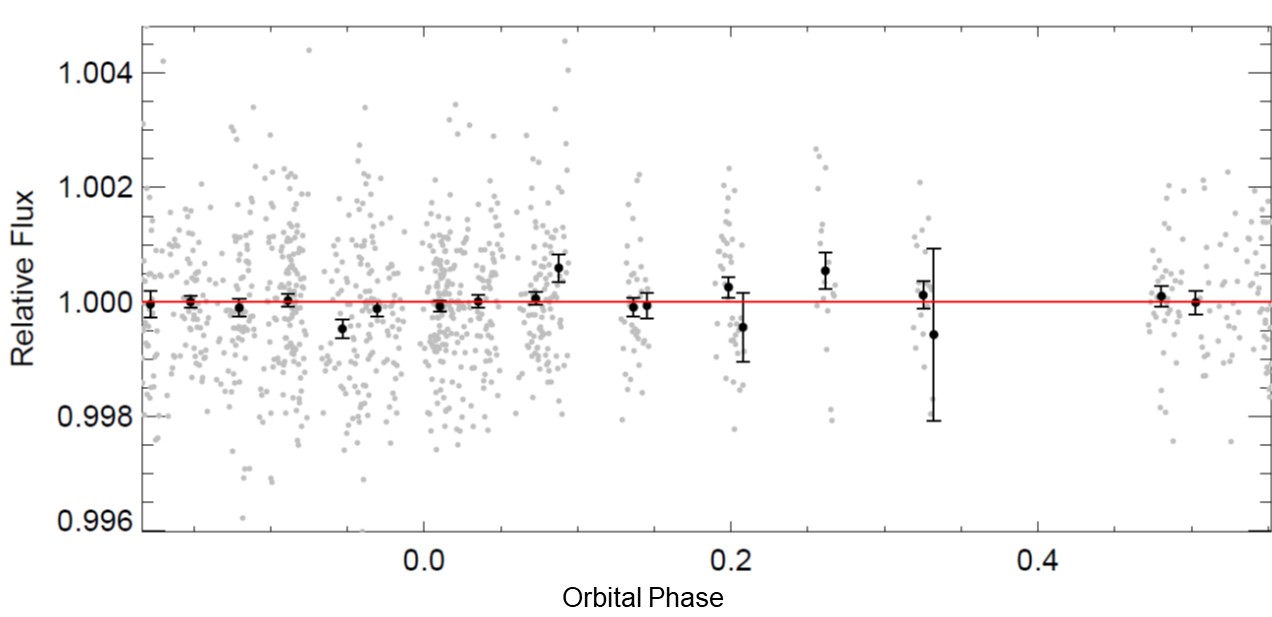}{0.75\textwidth}{c}
}\vspace{-5pt}
\caption{Phase-folded lightcurves for 55 Cancri e.  (a) Priors for $P$, $b$, and $T_0$ used; (b) Priors for $P$, $b$; (c) No transit model fit; phase-folded at the literature $P$ value.}
\label{fig:full_lc}
\end{figure*}

We then assessed our capability to detect 55 Cnc e transit in the ASTERIA data, without any prior knowledge on the planet (Figure~\ref{fig:full_lc}, bottom). To this end we used our detrended data as input to the \texttt{TLS} code \citep{Hippke2019} to compute a periodogram (Figure~\ref{fig:periodogram}. We identified the highest peak with an orbital period of 0.737 day, almost exactly the orbital period of 55 Cnc e. However this signal has a Signal Detection Efficiency \citep[SDE,][]{Kovacs2002} value of 4.9, which is not a firm detection (SDE$>$7). This finding is consistent with the MCMC fits conducted above.

To assess the planet search completeness of our 55 Cnc dataset, we conducted an injection-recovery test using the \texttt{allesfitter} code (G\"unther \& Tansu in prep). As the purpose of this step of the analysis was to determine the detection threshold at 55 Cnc e's ephemeris, we injected different planet sizes at the same $T_0$ and period. We find that we exceed a SDE of 7 \citep{Siverd2012} for planetary radii greater than 2.6~\Rearth  (vs.\ 55 Cnc e's $\sim1.9$~\Rearth).

\begin{deluxetable}{lcc}[h!]
\tablecaption{55 Cnc e transit parameters \label{tab:results}}
\tablewidth{0pt}
\tablehead{
\colhead{Parameter } & \colhead{Median \& 1-$\sigma$ credible interval} & \colhead{Source }} 
\startdata
Transit depth [ppm]             & $374 \pm 170$                 & This work    \\
Impact parameter $b$            & $0.40 \pm 0.03$               & prior \citep{Sulis2019} \\
Transit T0 [BJD]                & $8200.4343^{0.0026}_{0.0024}$ & prior \citep{Sulis2019}  \\
Period $P$ [days]               & $0.7365450 \pm 0.0000001$     & prior \citep{Sulis2019}   \\
\enddata 
\end{deluxetable}

\section{Summary and discussion} \label{sec:discuss}
The ASTERIA CubeSat observed 55 Cancri e, a known transiting super-Earth orbiting a Sun-like star.  The 55 Cnc data obtained with the ASTERIA CubeSat showed a marginal ($\sim$2.5$\sigma$) transit when fitted using priors from literature, but did not succeed in independently detecting the transit of 55 Cnc e, by a small margin. Our MCMC fits, completeness studies, and transit search demonstrate together that a signal is seen in ASTERIA data, but not at a level that is significant enough to claim independent detection without prior knowledge of the planet orbit and transit.  Additional photometric data on other stars was obtained during ASTERIA's prime mission and extended missions; results will be presented in future publications using the photometric reduction framework described here.

ASTERIA demonstrated subarcsecond pointing stability and $\pm$10 milliKelvin thermal control using passive cooling and active heating \citep{Smith2018, Pong2018}.  ASTERIA has matured these key technologies to enable high-precision photometry in a small package for future astrophysics CubeSat/smallsat missions and demonstrated the utility of small spacecraft for cutting-edge photometric measurements above the blurring effects of the Earth's atmosphere.

\subsection{Lessons learned}
ASTERIA's design was driven by technology demonstration requirements rather than science requirements.  As such, there are several modifications and improvements that would enhance the capabilities of next generation astrophysics CubeSat missions. 

\subsubsection{Orbit-transit phasing}
ASTERIA data showed excess systematic noise at timescales similar to the spacecraft orbit (90 minutes) and the duration of eclipse (20-30 minutes).  The duration of the 55 Cnc e transit (96 minutes) is similar to the duration of ASTERIA's orbit, causing `clumping' of data around the transit phase over short timescales; multiple observations of the transit did not lead to uniform filling of the lightcurve in phase space.  This effect led to a bias in the retrieved transit midpoint and depth (Section~\ref{sec:priors}).  Current and future photometry missions with non-continuous coverage may be subject to similar bias if astrophysical periods (orbit, transit) are similar to instrumental periods (spacecraft orbit, observation duty cycle).

\subsubsection{Detector capability}
First, a detector architecture where subarrays could be read out at different rates would significantly expand the set of target stars available for observation.  ASTERIA can only observe stars brighter than V=7 because there are not enough photons arriving from dimmer stars to reliably flip the first ADC bit in each 50 msec exposure.  If no signal is registered by the ADC, no amount of co-adding will produce a usable signal.  A CMOS imager and FPGA readout architecture capable of reading out guide star windows at 20 Hz for fine pointing control and science windows at a slower rate, allowing the collection of more photons per exposure, would allow a telescope with the same aperture size as ASTERIA to observe dimmer stars.

\subsubsection{Absolute time tagging}
Future CubeSat photometry or spectroscopy missions would also benefit from improved time-tagging of photometric data.  ASTERIA does not have a realtime clock or a functional GPS, so the flight computer clock resets to zero at every spacecraft reset.  Absolute time-tagging was not necessary for the technology demonstration goals of the mission, so ASTERIA was launched with a non-functional GPS unit.  The flight computer boot time in UTC must be calculated after each reset by comparing the UTC send time of a command with the logged receive time of the command in flight computer time, which is recorded as seconds from boot.  Uncertainty in the UTC-spacecraft clock correlation is $\geq$1 second because the ground data system records command send times with a precision of one second.  A working GPS unit would solve this problem, as would an onboard realtime clock that does not reset when the flight computer reboots.  One second time correlation precision is sufficient for the purposes of detecting transits with durations of many minutes to hours, but improved timing precision would open up additional applications, such as transit timing variation (TTV) measurement, for future missions.

\subsubsection{Pre-flight camera characterization}
The test campaign for ASTERIA's payload was designed to ensure that it would be capable of demonstrating the key technologies of fine pointing control and thermal control.  A future science-focused CubeSat photometry mission would benefit from pre-flight collection of bias, flat, and dark frames, and gain values, in multiple detector modes (full frame, windowed, co-added) over a range of flight-like temperatures.  These calibration data could be used for reducing the photometric data rather than collecting calibration data on-orbit, where illumination conditions are not as easily controlled.  The volume of available calibration data would also be much larger than what can be reasonably collected and downlinked from orbit.  See \citet[Chapter 5]{Krishnamurthy2020}

\subsubsection{Orbit selection}
ASTERIA's orbit is not optimal for photometric timeseries observations because there is no continuous viewing zone and the observing duty cycle is capped at $\sim$30\% due to the requirement that observations take place in eclipse.  A low Earth sun synchronous orbit would provide a stable thermal environment, near-continuous power generation, and longer continuous viewing.  The MOST space telescope \citep{Walker2003} used such an orbit and several upcoming CubeSat and smallsat missions will as well (SPARCS \citep{Shkolnik2018,Scowen2018},  CHEOPS \citep{Broeg2013}).  A low inclination orbit might be considered for future CubeSat space telescopes as well, as it provides continuous viewing of the celestial poles assuming sufficient baffling for stray light from the Sun and illuminated Earth.  A low inclination orbit also avoids SAA passages, which would be important for sensitive detectors and/or electronics.

See \citet{Smith2018} and \citet{Donner2018} for in-depth discussion of other lessons learned from ASTERIA development, integration and testing, and on-orbit operations.

\subsection{Next steps} \label{sec:nextsteps}
Orbit decay projections show that ASTERIA will remain in orbit until at least spring 2020.  Science operations ceased as of December 5, 2019 when contact with the spacecraft was lost.  During its extended missions, ASTERIA also served as a testbed for new flight software and navigation techniques \citep{Fesq2019}.

The technologies and techniques ASTERIA has demonstrated are enabling for future astrophysics CubeSats and smallsats, including the ExoplanetSat Constellation (described below).  A platform like ASTERIA with very stable pointing and thermal control could support a wide range of instruments, including photometers operating outside of the visible range (UV/IR) or low resolution spectrometers.  It should be possible to scale the ASTERIA platform up in volume to a 12U+ form factor in order to increase aperture size without significantly changing the hardware and algorithms used to achieve subarcsecond pointing stability.  ASTERIA was intended to be the first step toward a diverse fleet of small astrophysics missions, increasing access to space for the astrophysics community.

ASTERIA was a successful technology demonstration of a future constellation of up to dozens of satellites, dubbed the ExoplanetSat Constellation. Each satellite would share ASTERIA's precision pointing and thermal control capabilities, operate independently from the others, but may have different aperture sizes in order to reach down to fainter stars than ASTERIA's current capability. The primary motivation is the fact that if there is a transiting Earth size planet in an Earth-like orbit about the nearest, brightest (V$<$7) Sun-like stars, we currently have no way to discover them; current missions saturate on these bright stars. The ultimate goal for the constellation is to monitor dozens of the brightest sun-like stars, searching for transiting Earth-size planets in Earth-like (i.e., up to one year) orbits. Because the brightest sun-like stars are spread all across the sky, a single telescope will not do. Instead, each satellite would monitor a single sun-like star target of interest for as long as possible, before switching to another star, with targets only limited by the Sun, Earth and Moon constraints. To narrow down the approximately 3,000 target stars brighter than V=7, one would have to find a way to constrain the stellar inclinations and assume the planets orbit within about 10 degrees of the star’s equatorial plane. This would reduce the number of target stars from about 3000 to about 300 \citep{Beatty2010}, a much more tractable number of targets. The ExoplanetSat Constellation has a unique niche in context of existing and planned space transit surveys (Section \ref{sec:intro}), but is still in concept phase.

\acknowledgments
The authors thank the anonymous reviewer for their insightful and detailed comments which enhanced and improved the manuscript.
The authors recognize the contributions of the extended team that supported ASTERIA development, integration and test, and operations, including Len Day, Maria de Soria Santacruz-Pich, Carl Felten, Janan Ferdosi, Kristine Fong, Harrison Herzog, Jim Hofman, David Kessler, Roger Klemm, Jules Lee, Jason Munger, Lori Moore, Esha Murty, Chris Shelton, David Sternberg, Rob Sweet, Kerry Wahl, Jacqueline Weiler, Thomas Werne, Shannon Zareh, and Ansel Rothstein-Dowden.  We also recognize the JPL line organization and technical mentors for the expertise they provided throughout the project.

We also wish to recognize JPL program management, especially Sarah Gavit and Pat Beauchamp, who oversaw ASTERIA within the Engineering and Science Directorate at JPL.  We also thank Daniel Coulter and Leslie Livesay for their support.  

Finally, we would like to thank the DSS-17 ground station team at Morehead State University (MSU) in Kentucky. We acknowledge the outstanding efforts of the student operators, technical staff, and program management at Morehead State University, including Chloe Hart, Sarah Wilczewski, Alex Roberts, Maria Lemaster, Lacy Wallace, Rebecca Mikula, Bob Kroll, Michael Combs, and Benjamin Malphrus.  

We also thank Frank D. Lind and Mike Poirier of MIT Haystack observatory for their assistance in tracking ASTERIA during a communications anomaly.

Funding was provided by the JPL Phaeton program and by the Heising-Simons Foundation.

B.-O. D. acknowledges support from the Swiss National Science Foundation (PP00P2-163967).

The research was carried out in part at the Jet Propulsion Laboratory, California Institute of Technology, under a contract with the National Aeronautics and Space Administration (80NM0018D0004). 

This research has made use of the NASA Exoplanet Archive, which is operated by the California Institute of Technology, under contract with the National Aeronautics and Space Administration under the Exoplanet Exploration Program.
%

\vspace{5mm}


\software{astropy \citep{TheAstropyCollaboration2013}, astroquery \citep{Ginsburg2018}, photutils \citep{Bradley_2019_2533376}, pyBJD (https://github.com/tboudreaux/pyBJD), matplotlib \citep{Hunter2007}, IDL, STK, MATLAB}



\appendix
\startlongtable
\begin{deluxetable}{clrchc}
\tablecaption{55 Cancri observations used in the light curve shown in Figure \ref{fig:full_lc}. \label{tab:listobs}}
\tablewidth{0pt}
\tabletypesize{\scriptsize}
\tablehead{
\colhead{Observation} & \colhead{First frame} & \colhead{Last frame} & \colhead{Number of 1-minute}  & \nocolhead{Phase range} & \colhead{RMS}  \\
\colhead{name} & \colhead{[UTC]} & \colhead{[UTC]} & \colhead{integrations} &
 \colhead{} & \colhead{[ppm/minute]} 
}
\startdata
Tech Demo 14  & 2018-01-07 11:33:36.7 & 2018-01-07 11:52:36.5 & 20 & -0.063 -- -0.045 & 1123 \\
 (td14) & 2018-01-07 13:06:39.3 & 2018-01-07 13:25:39.1 & 20 & 0.025 -- 0.042 & \\
 & 2018-01-07 14:28:39.5 & 2018-01-07 14:47:39.3 & 19 & 0.102 -- 0.120 & \\
 \hline
Tech Demo 16  & 2018-01-10 10:26:35.5 & 2018-01-10 10:45:35.2 & 20 & -0.053 -- -0.035 & 986 \\
 (td16) & 2018-01-10 12:01:39.9 & 2018-01-10 12:20:39.7 & 19 & 0.036 -- 0.054 &\\
 & 2018-01-10 13:25:57.3 & 2018-01-10 13:44:57.1 & 20 & 0.116 -- 0.134 &\\
 \hline
Tech Demo 18  & 2018-01-12 10:10:29.5 & 2018-01-12 10:29:29.3 & 19 & -0.353 -- -0.335 & 1088 \\
 (td18) & 2018-01-12 11:43:03.9 & 2018-01-12 12:02:03.7 & 20 & -0.266 -- -0.248 & \\
 & 2018-01-12 13:15:37.5 & 2018-01-12 13:34:37.3 & 20 & -0.178 -- -0.161 & \\
 & 2018-01-12 14:54:42.8 & 2018-01-12 15:13:42.6 & 20 & -0.085 -- -0.067 & \\
 & 2018-01-12 16:27:47.8 & 2018-01-12 16:46:47.6 & 20 & 0.003 -- 0.021 & \\
 & 2018-01-12 17:57:21.4 & 2018-01-12 18:16:21.2 & 20 & 0.087 -- 0.105 & \\
 & 2018-01-12 19:36:16.6 & 2018-01-12 19:55:16.4 & 20 & 0.180 -- 0.198 & \\
 & 2018-01-12 21:04:31.3 & 2018-01-12 21:23:31.1 & 19 & 0.264 -- 0.282 & \\
 & 2018-01-12 22:39:02.9 & 2018-01-12 22:57:02.7 & 18 & 0.353 -- 0.370 & \\
 & 2018-01-13 00:10:37.6 & 2018-01-13 00:29:37.4 & 20 & 0.439 -- 0.457 & \\
 \hline
Tech Demo 23 & 2018-01-22 06:24:41.2 & 2018-01-22 06:44:41.0 & 21 & 0.011 -- 0.030 & 1016 \\
 (td23) & 2018-01-22 07:57:17.0 & 2018-01-22 08:17:16.7 & 21 & 0.098 -- 0.117 & \\
 & 2018-01-22 09:29:55.1 & 2018-01-22 09:49:54.9 & 21 & 0.186 -- 0.205 & \\
 & 2018-01-22 10:59:28.3 & 2018-01-22 11:19:28.1 & 21 & 0.270 -- 0.289 & \\
 \hline
Tech Demo 24 & 2018-01-22 18:44:23.7 & 2018-01-22 19:04:23.5 & 21 & -0.291 -- -0.273 & 998 \\
 (td24) & 2018-01-22 20:15:59.5 & 2018-01-22 20:35:59.3 & 21 & -0.205 -- -0.186 & \\
 & 2018-01-22 21:47:35.2 & 2018-01-22 22:07:35.0 & 20 & -0.119 -- -0.100 & \\
 & 2018-01-22 23:16:11.0 & 2018-01-22 23:36:10.7 & 20 & -0.035 -- -0.016 & \\
 \hline 
Science Observation 19 & 2018-04-13 20:44:02.5 & 2018-04-13 21:03:02.3 & 18 & -0.205 -- -0.187  & 798\\
(so19) & 2018-04-13 22:16:36.3 & 2018-04-13 22:35:36.2 & 20 & -0.118 -- -0.100 & \\
 & 2018-04-14 12:08:49.6 & 2018-04-14 12:27:49.4 & 20 & -0.333 -- -0.315 & 1193\\
 & 2018-04-14 13:41:18.5 & 2018-04-14 14:00:18.3 & 20 & -0.246 -- -0.228 & \\
 & 2018-04-14 15:13:46.3 & 2018-04-14 15:32:46.0 & 20 & -0.159 -- -0.141 & \\
 & 2018-04-14 18:18:44.6 & 2018-04-14 18:37:44.4 & 20 & 0.0158 -- 0.034 \\
 & 2018-04-15 09:42:29.4 & 2018-04-15 10:01:29.2 & 20 & -0.113 -- -0.095 & 1135\\
 & 2018-04-15 12:47:29.5 & 2018-04-15 13:06:29.3 & 20 & 0.061 -- 0.079 & \\
 & 2018-04-16 01:07:25.4 & 2018-04-16 01:26:25.2 & 20 & -0.241 -- -0.223 & 1429\\
 & 2018-04-16 02:39:50.4 & 2018-04-16 02:58:50.2 & 20 & -0.154 -- -0.136 & \\
 & 2018-04-16 05:44:46.9 & 2018-04-16 06:03:46.7 & 20 & 0.020 -- 0.038 & \\
 & 2018-04-16 07:17:15.2 & 2018-04-16 07:36:15.0 & 20 & 0.108 -- 0.125 & \\
 \hline
Science Observation 20 & 2018-04-22 00:28:32.3 & 2018-04-22 00:43:32.1 & 16 & -0.132 -- -0.117 & 1956 \\
(so20) & 2018-04-22 02:01:01.9 & 2018-04-22 02:16:01.8 & 16 & -0.044 -- -0.030 & \\
 & 2018-04-22 03:33:32.6 & 2018-04-22 03:48:32.5 & 15 & 0.043 -- 0.057 & \\
 & 2018-04-22 05:06:02.7 & 2018-04-22 05:21:02.5 & 16 & 0.130 -- 0.144 & \\
 \hline
Science Observation 24 & 2018-05-12 13:29:38.9 & 2018-05-12 13:48:38.6 & 19 & -0.241 -- -0.223 & (not used) \\
(so24) & 2018-05-12 16:38:23.5 & 2018-05-12 16:57:23.3 & 20 & -0.063 -- -0.045 & (not used)\\
 & 2018-05-12 18:04:49.5 & 2018-05-12 18:23:49.3 & 18 & 0.018 -- 0.036 & 639 \\
 \hline
Science Observation 25 & 2018-05-13 09:32:12.7 & 2018-05-13 09:51:12.5 & 19 & -0.107 -- -0.089 & 1031 \\
(so25) & 2018-05-13 11:08:36.4 & 2018-05-13 11:27:36.2 & 20 & -0.016 -- 0.002 & \\
 & 2018-05-13 12:33:59.2 & 2018-05-13 12:52:59.0 & 19 & 0.064 -- 0.082 & \\
 & 2018-05-14 02:26:01.8 & 2018-05-14 02:45:01.6 & 19 & -0.151 -- -0.133 & 637\\
 & 2018-05-14 19:26:05.3 & 2018-05-14 19:45:05.1 & 20 & -0.190 -- -0.172 & 1366\\
 & 2018-05-14 20:58:34.1 & 2018-05-14 21:17:33.9 & 20 & -0.102 -- -0.085 & \\
 & 2018-05-15 00:00:25.9 & 2018-05-15 00:19:25.7 & 20 & 0.069 -- 0.087 & \\
 \hline
Science Observation 26 & 2018-05-15 12:20:22.2 & 2018-05-15 12:41:22.0 & 22 & -0.233 -- -0.214 & 1487 \\
(so26) & 2018-05-15 15:25:19.8 & 2018-05-15 15:46:19.6 & 20 & -0.059 -- -0.039 & \\
 & 2018-05-15 16:57:46.7 & 2018-05-15 17:18:46.4 & 22 & 0.028 -- 0.048 & \\
 \hline
Science Observation 28 & 2018-05-26 07:21:00.9 & 2018-05-26 07:35:00.7 & 15 & &  (not used)\\
(so28) & 2018-05-26 08:53:20.2 & 2018-05-26 09:07:20.1 & 15 & & (not used)\\
 & 2018-05-26 10:25:40.1 & 2018-05-26 09:07:20.1 & 15 & & (not used)\\
 & 2018-05-27 09:29:45.0 & 2018-05-27 09:34:45.0 & 6 & & (not used)\\
 & 2018-05-27 12:34:25.9 & 2018-05-27 12:39:25.8 & 6 & & (not used)\\
 & 2018-05-28 20:53:39.3 & 2018-05-28 20:58:39.2 & 6 & & 810\\
 & 2018-05-28 22:26:00.8 & 2018-05-28 22:31:00.7 & 6 & & \\
 & 2018-05-28 23:58:21.4 & 2018-05-29 00:02:21.4 & 5 & & \\
 \hline
Science Observation 30 & 2018-06-08 18:36:29.9 & 2018-06-08 18:43:29.9 & 8 & & 1535\\
(so30) & 2018-06-08 20:08:52.2 & 2018-06-08 20:15:52.1 & 8 & & \\
 & 2018-06-08 21:42:13.2 & 2018-06-08 21:48:13.2 & 7 & & \\
 & 2018-06-08 23:13:35.9 & 2018-06-08T 3:20:35.8 & 8 & & \\
 & 2018-06-10 06:00:53.8 & 2018-06-10 06:07:53.7 & 8 & & 1054\\
 & 2018-06-10 07:33:17.4 & 2018-06-10 07:40:17.4 & 8 & & \\
 & 2018-06-10 09:05:37.9 & 2018-06-10 09:12:37.8 & 8 & & \\
 & 2018-06-10 10:38:01.3 & 2018-06-10 10:45:01.2 & 8 & & \\
 \enddata
\tablecomments{Observations designated `Tech Demo' or `td' took place during the initial phase of the mission when observations were focused on addressing ASTERIA's technology demonstration goals.  After the technology demonstration objectives were completed, the naming scheme changed to 'Science Observation' or 'so' to indicate the change in focus from technology demonstration to collection of science data.  Lightcurve and image data will be available at \url{https://exoplanetarchive.ipac.caltech.edu/docs/ASTERIAMission.html}}
\end{deluxetable}





\bibliography{references_55Cnc}




\end{document}